    \documentclass{IEEEtran}
  \IEEEoverridecommandlockouts
  \usepackage{cite}
  \usepackage{amsmath,amssymb,amsfonts}
  \usepackage{algorithmic}
  \usepackage{graphicx}
  \usepackage{textcomp}
  \usepackage{xcolor}
  \usepackage{setspace}
  \usepackage{enumitem}
  \def\BibTeX{{\rm B\kern-.05em{\sc i\kern-.025em b}\kern-.08em
  		T\kern-.1667em\lower.7ex\hbox{E}\kern-.125emX}}
  \usepackage{subfigure}
  \newtheorem{my_theorem}{Theorem}
  
  \newtheorem{my_lemma}{Lemma}
  \newtheorem{my_corollary}{Corollary}
  \newtheorem{my_proposition}{Proposition}
  
  \usepackage{hhline}
\title{Performance of Dual-Hop  Relaying for OWC System Over  Foggy 	Channel with Pointing Errors and Atmospheric Turbulence}
\author{Ziyaur Rahman,~\IEEEmembership{Gradute Student Member,~IEEE}, Tejas Nimish Shah, S.~M.~ Zafaruddin,~\IEEEmembership{Senior Member,~IEEE,} and
	V.~K.~ Chaubey,~\IEEEmembership{Senior Member,~IEEE}
	\thanks{This work was supported in part by the Start-up Research Grant, Science and Engineering Research Board (SERB), Department of Science and Technology (DST), India under Grant SRG/2019/002345.
		
		Ziyaur Rahman (p20170416@pilani.bits-pilani.ac.in), Tejas Nimish Shah (f20170024@pilani.bits-pilani.ac.in), S.~M.~Zafaruddin (syed.zafaruddin@pilani.bits-pilani.ac.in), and V.~K.~Chaubey (vkc@pilani.bits-pilani.ac.in)  are  with  the Department of Electrical and Electronics Engineering, Birla Institute of Technology and Science, Pilani, Pilani-333031, Rajasthan, India.}
	
	\thanks{}}

\begin{document}
	\maketitle
	\begin{abstract}\label{sec:abstract}
Optical wireless communication (OWC) over atmospheric turbulence and pointing errors  is a well-studied topic. Still, there is  limited research on  signal fading due to  random fog in an  outdoor environment for terrestrial wireless communications. In this paper, we analyze the performance of a decode-and-forward (DF) relaying  under the combined effect of random fog,  pointing errors, and atmospheric turbulence with a negligible line-of-sight (LOS) direct link.  We consider a generalized  model for the end-to-end channel  with independent and  not identically distributed (i.ni.d.) pointing errors, random fog with Gamma distributed attenuation coefficient, double generalized gamma (DGG) atmospheric turbulence,    and asymmetrical distance between the source and destination.  We develop density and distribution functions of  signal-to-noise ratio (SNR) under the combined effect of random fog, pointing errors, and atmospheric turbulence (FPT) channel  and distribution function for the combined channel with random fog and pointing errors (FP). Using the derived statistical results,  we present analytical expressions of the outage probability, average SNR, ergodic rate, and average bit error rate (BER) for both FP and FPT channels in terms of OWC system parameters. We also develop  simplified and asymptotic performance analysis to provide insight on the system behavior analytically under various practically relevant scenarios.   We demonstrate the mutual effects of channel impairments and pointing errors on the OWC performance, and show that the  relaying system  provides significant performance improvement compared with the direct transmissions,  especially when  pointing errors and fog becomes more pronounced.
\end{abstract}

	
\begin{table*}[tp]
	
	\caption{Literature on FSO Systems with Random Fog  }
	\label{table:diversity_order}
	\centering
	\begin{tabular}{ c c c c p{8cm} c}
		\hline
		\hline
		Reference   & System Model &Pointing Errors & Turbulence & Performance Metrics Analysis\\
		\hline
	  [26] &  Direct link, Single-aperture& No  &No & BER (numerical), Channel Capacity (numerical) \\  \hline
	  [27]& Direct link, Single-aperture &No &  No &Outage probability, SNR, Channel Capacity (numerical), BER (numerical)\\ \hline
	  [28] &  Direct link, Multi-aperture& No & No & Outage probability, SNR, Channel Capacity\\ \hline
	  [30] &  Multi-hop, Single-aperture & Yes & No & Outage probability \\ \hline
	   [31] &  Direct link, Multi-aperture & Yes & No & Outage probability, SNR, Channel Capacity \\ \hline
	  [Proposed] & Dual-hop, Single-aperture &Yes &Yes& Outage probability, SNR, Channel Capacity, BER\\
	  \hline
	  \hline
	
	\end{tabular}

\end{table*}

	\section{Introduction}\label{sec:introduction}
Optical wireless communication (OWC) is a potential technology that transmits data through an unguided atmospheric channel  in the unlicensed optical spectrum \cite{Khalighi2014, Bloom2003, Kedar2004}.  However, signal transmission at a small wavelength encounters  different channel impairments such as  atmospheric turbulence,  pointing errors, and fog. The atmospheric turbulence  is caused by  the scintillation effect of light propagation whereas the  pointing errors (i.e.,  misalignment between the transmitter and receiver) happens due to the  dynamic wind loads, weak earthquakes, and thermal expansion \cite{Farid2007, Vavoulas2012}.  The impact of foggy conditions on OWC systems depends on the intensity of fog ranging between light, medium, and dense \cite{Kruse1962,Kim2001}.  Although turbulence and fog may not co-exist since both are  inversely correlated with each other \cite{Bushuev2006, Farid2007},  the effect of turbulence  can not be ignored in light foggy conditions. The combined effect of atmospheric turbulence, pointing errors, and fog has a  detrimental  effect on  the signal quality and presents a major challenge  in the  OWC deployment in outdoor environments.


The use of relaying  has  been extensively studied to improve the performance of OWC systems under the effect of turbulence and/or pointing errors \cite{Safari2008, Aghajanzadeh2011, multi_hop_turb2015, Chatzidiamantis2013,parallel_fso2015,  multi_fso2013, parallel_multi_fso2016, Karimi2011, Kazemlou2011, Bayaki2012, Kashani2012, Trinh2015,Yang2014_relay,  dual_hop_turb2017, Dabiri2018, Huang2018}. In the aforementioned and related research, the statistical  effect of foggy channels combined with pointing errors and atmospheric turbulence has not been considered. Recent studies confirm that the signal attenuation  in the fog is not deterministic but follows a probabilistic model  \cite{Khan2009, Esmail2016_Photonics,Esmail2017_Access}.  In \cite{Esmail2016_Photonics}, the authors developed Johnson SB based probability distribution function (PDF) as a model for the random fog channel. They studied numerically the system performance in terms of average bit error rate (BER) and channel capacity. Considering the intractability of the Johnson SB for performance analysis,  the authors in \cite{Esmail2017_Access} proposed Gamma distribution for the signal attenuation in foggy weather and evaluated average signal-to-noise ratio (SNR), ergodic rate, outage probability, and BER. In our recent paper \cite {Rahman2020}, we  proposed a multi-aperture OWC system to mitigate the effect of fog. The authors in \cite{Schimmel2018} presented ultrashort high-intensity laser filaments for high-bit-rate transmissions over the fog.  These studies show that  the OWC performance is significantly limited in dense fog but can provide  acceptable performance in the light fog over shorter links.  However, combining the effect of pointing errors with fog shows high degradation in performance even in light foggy conditions\cite{Esmail2017_Photonics, rahman2020cl}. Specifically, the authors in \cite{Esmail2017_Photonics} have considered three techniques to  mitigate this effect: multi-hop relay systems using decode-and-forward (DF), active laser selection, and parallel radio frequency/free space optical (RF/FSO)  link.  Laser selection and hybrid transmission techniques require feedback from the receiver to the transmitter, thereby increasing the overhead.  On the other hand,  multi-hop relaying requires channel state information (CSI) at each relay to decode the signal, which can be hard in practice.  In \cite{Esmail2017_Photonics}, the outage probability (which requires direct application of the cumulative distribution function (CDF) of the channel) of a multi-hop FSO system under the combined effect of fog and pointing errors has been analyzed. However, there is no result available in the literature for other performance metrics such as average SNR, ergodic rate, and average BER even for  dual-hop OWC system under the random foggy channel with pointing errors. It is quite involved to derive closed-form expressions of the relay-assisted system since   distribution  functions of the resultant fading channel  consists of an incomplete gamma function with a logarithmic argument. Analyzing the system performance using different performance metrics such as average SNR, ergodic rate, and average BER is desirable for  efficient deployment of OWC systems under the combined effect of fog and pointing errors. Most importantly, the atmospheric turbulence has been ignored in the existing literature since deriving the PDF and CDF of the combined channel is complicated and requires novel approaches. It should be mentioned that the atmospheric turbulence can be neglected for shorter links but ignoring the turbulence for longer links  may underestimate/overestimate the performance of the OWC system.  To the best of the authors' knowledge,  there are no analyses available for  average SNR, ergodic rate, BER, and outage probability for the relay-assisted OWC system under the statistical effect of random fog, pointing errors, and atmospheric turbulence. In Table \ref{table:diversity_order},  we summarize the reported research in the literature on OWC systems under random fog.

In this paper,  we  analyze the end-to-end performance of a relay-assisted OWC system under the combined effect of  fog, pointing errors, and atmospheric turbulence (termed as the FPT channel) by considering a single  DF relaying with  no direct transmission to the destination. The major contributions of the paper are as follows:\vspace{-0.1mm}
\begin{itemize}[leftmargin=*]
\item We consider a generalized  model for the end-to-end channel of the relay-assisted system with independent and  not identically distributed (i.ni.d.) pointing errors, random fog with Gamma distributed attenuation coefficient, double generalized gamma (DGG) atmospheric turbulence, and asymmetrical distance between the source and destination. 

We also analyze the considered system under the effect of random fog and pointing errors (termed as the FP channel) with  negligible atmospheric turbulence for shorter OWC links.  
 
\item We derive a novel PDF of the SNR  of the OWC link for the FPT channel in terms of a single Meijer's-G function, which resulted into more elegant performance analysis.
	We also simplify the existing CDF of the FP channel in terms of a single incomplete gamma function to derived closed-form  expressions of the OWC system.
	\item We use the derived statistical results to develop  analytical expressions of the outage probability,  average SNR, ergodic rate, and average BER for both FP and FPT channels  with single-variate Fox-H function for the FPT channel  and standard mathematical functions for the FP channel.
	\item We  present asymptotic analysis in the high SNR 	regime for outage probability and average BER and derive diversity order depicting 	the impact of system and channel  parameters on the performance of the 	considered system.
	\item We use numerical and simulation analysis to show that the dual-hop relaying can mitigate fog,  pointing errors,  and turbulence-induced fading for high-speed OWC links. We also demonstrate  that there is a significant gap in the performance using the existing visibility range based path-gain computation as compared to the probabilistic modeling of random  fog.

\end{itemize}

\subsection{Related Research}

Traditionally, signal attenuation due to the fog was assumed to be deterministic and quantified using a visibility range, for example, less attenuation in light fog and more in the dense fog \cite{Kruse1962,Kim2001,Kedar2003, Naboulsi2004, Awan2008, Esmail2016_ICC,Esmail2016_Photonics, Berenguer2018}. Kruse model in \cite{Kruse1962} is based on experimental data, whereas Kim \cite{Kim2001} used Mie scattering theory to predict the signal attenuation. The authors in \cite{Naboulsi2004} \cite{Awan2008} updated the earlier models considering  modified Gamma distribution for the particle size of fog. The authors in \cite{Esmail2016_ICC} developed  a power delay model for the attenuation coefficient based on extensive measurement data. On the other hand, there are  quite a few  statistical models for the atmospheric turbulence, for example, log normal \cite{Zhu2002}, exponentiated Weibull (EW) \cite{Barrios2012}, Gamma-Gamma  \cite{Andrews20015},  Mal\'aga \cite{malaga2011},  and  ${\mathcal{F}}$-distribution \cite{Peppas2020}.

 Recently,  \cite{Kashani2015}  proposed  the DGG distribution model for  atmospheric turbulence. It is based on the theory of  doubly stochastic scintillation, where irradiance fluctuations are expressed   as the product of large-scale and small-scale fluctuations each following the generalized Gamma distribution. The DGG model can be used to model accurately different propagation conditions and it is versatile to  include several  statistical models for atmospheric turbulence as special cases. The authors in \cite{AlQuwaiee2015, Ashrafzadeh2020} analyzed the performance  of the OWC over DGG atmospheric turbulence without the consideration of random fog.   It should be noted that the model of pointing errors \cite{Farid2007} is used widely,  assuming independent identical distributed Gaussian  for the elevation and the horizontal displacement.

Relay-assisted communication is a potential technique to deal with the channel fading in wireless systems \cite{Nosratinia2004, Li2012}. Here, a single relay  or many intermediate nodes   can assist data transmission between a single source and destination. In particular, for OWC systems,  there is a vast literature on the relaying  using amplify-and-forward (AF) and DF protocols in  \cite{Safari2008, Aghajanzadeh2011, multi_hop_turb2015, Chatzidiamantis2013,parallel_fso2015,  multi_fso2013, parallel_multi_fso2016,AlQuwaiee2015}, all-optical relaying in  \cite{Karimi2011, Kazemlou2011, Bayaki2012, Kashani2012, Trinh2015,Yang2014_relay,  dual_hop_turb2017, Dabiri2018, Huang2018}, and relaying for hybrid  RF/FSO  systems in  \cite{Lee2011, Ansari2013, Samimi2013, assym_rf_fso2015, series_hybrid_m_channel2015, dual_hop_rf_fso_turb2016, Bag2018,Zhang2020}. In the seminal work \cite{Safari2008},  multi-hop and cooperative relaying  using AF and DF protocols  have been considered for an aggregated channel model which takes into account both path-gain and turbulence-induced log-normal fading. The authors in \cite{Aghajanzadeh2011} investigated a multi-hop relaying to mitigate the effect of fading in  FSO systems over log-normal atmospheric turbulence channels.  The end-to-end performance in terms of outage probability, the average BER, and the ergodic capacity for a multi-hop relaying with AF and DF protocols under the combined effect of Gamma-Gamma turbulence and pointing errors  have been investigated in \cite{multi_hop_turb2015}. Although the complex multi-hop relaying can provide a better performance, a dual-hop relay system (with no direct link between the source and the destination) that selects a single relay opportunistically is considered in \cite{Chatzidiamantis2013}. The authors in \cite{parallel_fso2015} studied the information-theoretic performance of parallel relaying for FSO communications over Gamma-Gamma fading channels  with a single relay  but with a line-of-sight link between the source and the destination.  An optimal relay placement
scheme for serial and parallel relaying along with a diversity gain analysis has been considered  in \cite{multi_fso2013}.  The authors in \cite{parallel_multi_fso2016} have considered the inter-relay cooperation on the outage probability and diversity order performance of the DF cooperative FSO communication systems.

An all-optical relaying scheme is  efficient  since signals are processed in the optical domain  without requiring  optical-to-electrical and electrical-to-optical conversions \cite{Karimi2011, Kazemlou2011, Bayaki2012, Kashani2012, Trinh2015,Yang2014_relay,  dual_hop_turb2017, Dabiri2018, Huang2018}. The references \cite{Yang2014_relay, dual_hop_turb2017} provided analytical expressions for the outage probability, average BER, and ergodic capacity over strong atmospheric turbulence  channels with misalignment-induced pointing errors by considering a single optical AF  relay with fixed and variable gain.  Hybrid RF/FSO systems, where relays act as an interface between RF and optical links, have been studied  \cite{Lee2011, Ansari2013, Samimi2013, assym_rf_fso2015, series_hybrid_m_channel2015, dual_hop_rf_fso_turb2016, Bag2018,Zhang2020}.   A dual-hop relay system over the asymmetric links has been considered for both RF and FSO environments in \cite{Lee2011} and derived exact expression for outage probability.  The BER performance and the capacity analysis of an AF-based dual-hop mixed RF–FSO is presented in \cite{Ansari2013}, where the RF links are Rayleigh distributed and FSO links are characterized by the Gamma–Gamma distributed turbulence and pointing errors.    Considering a dual-hop transmission with a single-relay and ignoring the direct transmission, the authors in  \cite{assym_rf_fso2015, series_hybrid_m_channel2015, dual_hop_rf_fso_turb2016} have analyzed the OWC performance under the  turbulence and pointing errors.    The authors in \cite{Bag2018} have considered an FSO/RF-FSO link adaptation scheme for hybrid FSO systems and analyzed  different performance metrics like outage probability, average BER, and ergodic rate. Recently, a hybrid dual-hop relaying with mmWave  and FSO scheme  is studied in \cite {Zhang2020}. 
 
\subsection{Notations and Organization}
We list the main notations  in Table \ref{table:notation_parameters}. This paper is organized as follows: In Section II, we discuss the relay assisted OWC system model. In Section III, we analyze the OWC system performance by deriving closed-form expressions for the outage probability, average SNR, ergodic rate, and average BER. In Section IV, the simulation results of the proposed system are presented. Finally, in Section V, we provide conclusions. 

\begin{table}[tp]
	\caption{The List of Main  Notations}
	\label{table:notation_parameters}
	\centering
	\begin{tabular}{|c|p{5cm}|c|}
		\hline
		 $(\cdot)_1$ & Notation for the first link \\ \hline  $(\cdot)_2$ & Notation for the second link\\ 
		 \hline
		$y$& Received signal \\ \hline	$h$ & Random channel state\\
		\hline
		$R$ & Detector responsivity\\	
		
		\hline
		$x$ & Transmit signal intensity \\ \hline 		$w$ & AWGN  \\
		\hline
		$P_t$ & Transmit power \\ \hline 
		$d_1$ & Distance: transmitter to relay \\ \hline 
		$d_2$ & Distance: relay and destination\\ \hline 
		$d$ & $d_1+d_2$ \\ \hline $k$ & Shape parameter of foggy channel\\
		\hline
		$\beta$ & Scale parameter of foggy channel \\ \hline 
		$P_{\rm out}$ & Outage probability\\ \hline
		$\gamma_0$ &  SNR without fading\\\hline
		$\gamma$  & SNR with fading \\ \hline 	
		
			$\bar{\gamma}$ & Average SNR\\
		\hline
		
				$\bar{\eta}$ & Ergodic rate \\ \hline	
				$\bar{P_e}$ & Average BER \\ \hline	
							$\Gamma(a)$ & $\int\limits_{0}^{\infty}t^{a-1} e^{-t}dt$ \\ \hline
				$\Gamma(a,t)$ & $\int_{t}^{\infty}s^{a-1}e^{-s}ds$\\ \hline 	$Q(\gamma)$ & $\frac{1}{\sqrt{2 \pi}}\int_{\gamma}^{\infty}e^{-\frac{u^2}{2}}du$\\ \hline
 $\rm{erf}(\gamma)$ &$\frac{2}{\sqrt{\pi}}\int_{0}^{\gamma}e^{-u^2}du$\\ \hline 
${}_1F_{1}(a;b;z)$ & $\sum_{k=0}^\infty \frac{\Gamma(a+k)\Gamma(b)}{\Gamma(a)\Gamma(b+k)}\frac{z^k}{k!}$ \\ \hline
	$\psi^{(0)}(a)$& $\frac{d}{da} \ln \Gamma(a)$ \\ \hline
	$G_{p,q}^{m,n}(.|.)$ & Meijer-G function \\ \hline
		$H_{p,q}^{m,n}(.|.)$ & Fox-H function \\ \hline
		\end{tabular}	
\end{table}
\begin{figure}[tp]	
	\centering
	\includegraphics[scale=0.3]{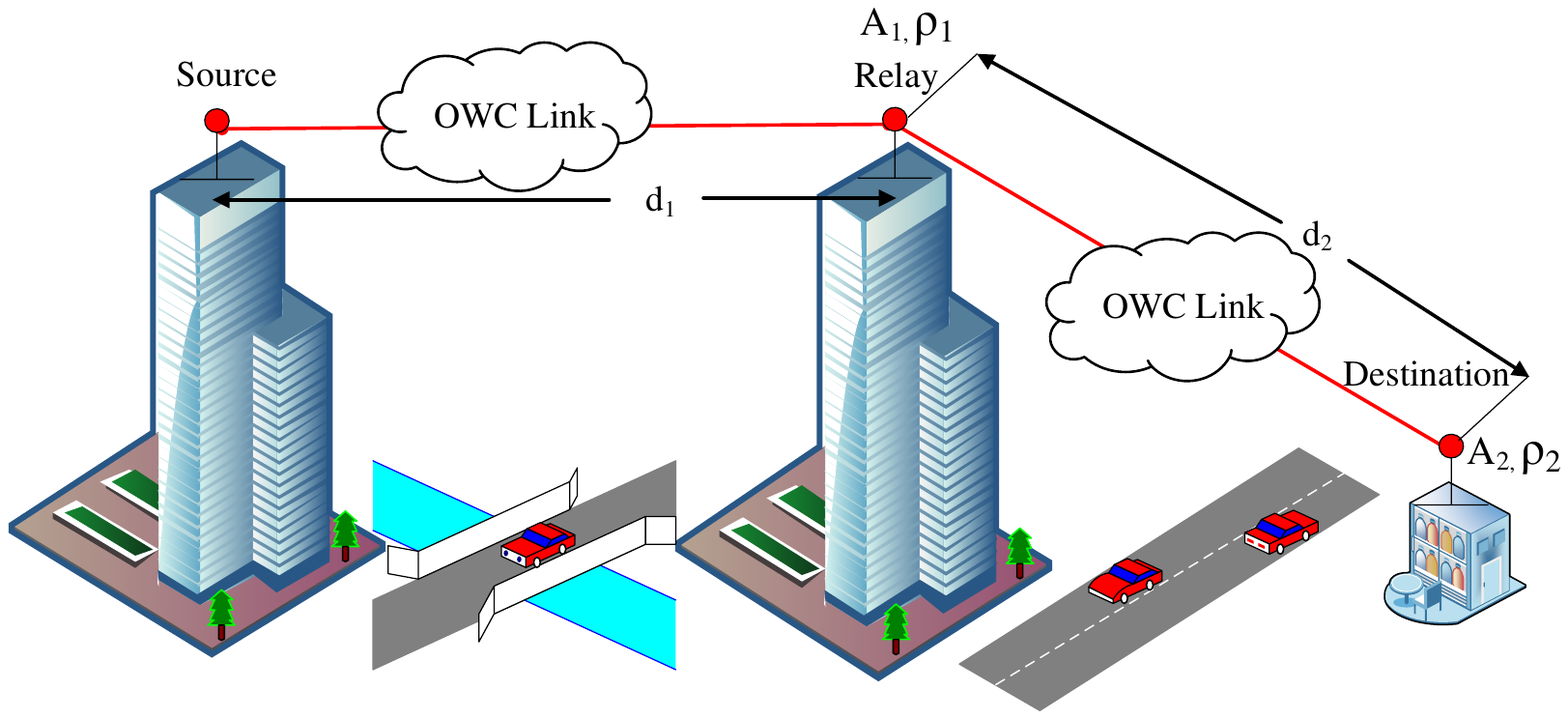}
	\vspace{-2cm}
	\caption{Relay assisted OWC system.}
	\label{system_model}
	
\end{figure}
\section{System Model}\label{sec:system_model}
We consider an OWC system using intensity modulation/direct detection (IM/DD). It consists of a single-aperture transceiver system, with a negligible line-of-sight (LOS) direct link for the FPT channel under the combined effect of  fog, pointing errors, and atmospheric turbulence, as shown in Fig~\ref{system_model}.
The signal $y_i$ at the $i$-th receiver aperture is given as 
\begin{eqnarray}
y_i = h_{fi} h_{pi} h_{ti}R_ix + w_i,
\label{eq:received_signal}
\end{eqnarray} 
where $x$ is the transmitted signal, $R_i$ represents the detector responsivity (in amperes per watt), and $w_i$ represents an additive white Gaussian noise (AWGN) with variance $\sigma^2_{w_i}$. The terms $h_{fi}$, $h_{pi}$, and $h_{ti}$ are the random states of the foggy channel, pointing errors, and atmospheric turbulence induced fading, respectively of the $i$-th link.

We use the PDF of fog as given in \cite{Esmail2017_Access}:
\begin{eqnarray}
f_{h_{fi}}(h_{f}) = \frac{z_i^{k_i}}{\Gamma(k_i)}\left(\ln\frac{1}{h_{f}}\right)^{k_i-1} h_{f}^{z_i-1},
\label{eq:fh}
\end{eqnarray}
where $0<h_{f}\leq 1$, $z_i=4.343/\beta_{i}^{\rm fog} d_i$, $k_i>0$ is the shape parameter and $\beta_{i}^{\rm fog}>0$ is the scale parameter. It is noted that different pairs of $k_i, \beta_i$  determine the severity of the foggy channel such as  $\{k_i=2.32, \beta_i^{\rm fog}=13.12\}$,  $\{k_i=5.49, \beta_i^{\rm fog}=12.06\}$, $\{k_i=6.0, \beta_i^{\rm fog}=23.06\}$ for light, moderate and thick foggy conditions, respectively. The PDF of pointing errors fading $h_{pi}$ is given in \cite{Farid2007}:
\begin{eqnarray}
f_{h_{pi}}(h_p) = \frac{\rho_i^2}{A_{i}^{\rho_i^2}}h_{p}^{\rho_i^{2}-1},0 \leq h_{p} \leq A_i,
\label{eq:pdf_hp}
\end{eqnarray}
where $A_i=\mbox{erf}(\upsilon_i)^2$ with $\upsilon_i=\sqrt{\pi/2}\ a_i/\omega_{zi}$ and $\omega_{zi}$ is the beam width,	and $\rho_i = {\frac{\omega_{{zi}_{\rm eq}}}{2 \sigma_{s_i}}}$ with  $\omega_{{zi}_{\rm eq}}$ as the equivalent beam width at the receiver and $\sigma^2_{si}$ as the variance of pointing errors displacement characterized by the horizontal sway and elevation \cite{Farid2007}. Finally,  PDF of the random atmospheric turbulence with DGG distribution combined with zero-bore sight pointing errors  \cite{AlQuwaiee2015}:
\begin{eqnarray}\label{eq:pdf_ha_dgg}
&	f_{h_{tpi}}(h_{tp}) = \frac{\rho_i^{2}\sigma_i^{\beta_i-\frac{1}{2}}\lambda_i^{\varphi_i-\frac{1}{2}} (2\pi)^{1-\frac{\lambda_i+\sigma_i}{2}}}{\Gamma(\beta_i)\Gamma(\varphi_i)h_{tp}}  G_{\lambda_i+\sigma_i+1,1}^{0,\lambda_i+\sigma_i+1}\nonumber \\& \bigg[\begin{array}{c}\frac{\lambda_i^{\lambda_i}\sigma_i^{\sigma_i}\Omega_i^{\sigma_i}\Xi_i^{\lambda_i}}{\beta_i^{\sigma_i}\varphi_i^{\lambda_i}} \big(\frac{A_{i}}{h_{tp}}\big)^{\phi_i\lambda_i}\end{array} \bigg\vert \begin{array}{c} \mu_i \\ \nu_i\end{array}\bigg],~0\leq h_{tp} < \infty 
\end{eqnarray}
where $\mu_i=1-\frac{\rho_i^2}{\phi_i\lambda_i},\Delta(\sigma_i:1-\beta_i),\Delta(\lambda_i:1-\varphi_i)$, $\nu_i=\frac{-\rho_i^2}{\phi_i\lambda_i}$, and $\Delta(x:y)\triangleq\frac{y}{x},\frac{y+1}{x},\cdots,\frac{y+x-1}{x}$. The sets $(\alpha_i,\beta_i,\Omega_i)$, $(\phi_i,\varphi_i,\Xi_i)$ are DGG fading parameters and $\rho_{i}$, $A_{i}$ are pointing error parameters.

We denote by $d_1$ the distance between the source and relay and $d_2$ the distance between the relay and  destination.  For the case of relayed transmission using the DF protocol, the expressions for signals received at the relay and destination when   $x$ is the transmitted signal:
\begin{eqnarray}
y_r = h_{f1} h_{p1}h_{t1}R_1x + w_1
\label{eq:received_signal_at_relay}
\end{eqnarray}
\begin{eqnarray}
y_d = h_{f2} h_{p2}h_{t2}R_2x + w_2
\label{eq:received_signal_at_destination}
\end{eqnarray}
where $h_{f1}$, $h_{p1}$, $h_{t1}$ and $h_{f2}$, $h_{p2}$, $h_{t2}$ are random fog, pointing errors, and atmospheric turbulence channel states between source-relay and relay-destination, respectively, each having $w_1$ and $w_2$ as AWGNs. Note that $h_{1} = h_{f1}h_{p1} h_{t1}$ is the combined channel between the source and the relay, and $h_{2} = h_{f2}h_{p2} h_{t2}$ is the combined channel between the relay and destination.

We use a general relaying scenario  $d_2> d_1$  with different values of pointing errors parameters for both  links giving an i.ni.d fading model for the FPT channel.  To simplify the analysis, we assume  random fog to be  i.i.d. for the the FP channel since the foggy weather may affect both the links independently generated from the same probabilistic model. As a special case, we also consider that  the relay is situated at the midway  between the source and destination (i.e., $d_1=d_2$) and that the channel parameters are same for both links (i.e., i.i.d. condition). 

\section{Performance Analysis}\label{sec:performance_analysis}
In this section, we analyze the performance of a relay-assisted system. First, we present exact expressions for the PDF and CDF of the  SNR for the FPT channel  under the combined effect fog, pointing errors, and atmospheric turbulence. We also provide a 	simplified expression for the FP channel under the   effect fog and pointing errors with negligble atmospheric turbulence.  Next, we use the derived statistical results to analyze the outage probability, average SNR, ergodic rate, and average BER performance of the OWC system. The derived expressions show the system behavior in a relay-assisted environment under the combined effect of channel impairments that can help the network operator for an efficient design of relay-assisted OWC links.

\subsection{Statistical Results}
We define $\gamma_i^{\rm FPT}=\gamma_0 |h_i|^2$ as the SNR for the FPT channel, where $h_i = h_{fi}h_{pi} h_{ti}$, $\gamma_0=2P^2_tR_i^2/\sigma^2_{wi} $ for $i=1, i= 2$,  and $P_t$ is the average optical transmitted power. In the following Theorem, we provide PDF and CDF for the FPT channel under the combined effect of the random fog,  pointing errors, and atmospheric turbulence:
\begin{my_theorem}[PDF and CDF for FPT]\label{snr_combine}
	The PDF and CDF of SNR for the FPT channel  under the combined effect of random fog and atmospheric turbulence with pointing errors for the single OWC link is given as
	\begin{eqnarray}
		&{f^{\rm FPT}_{\gamma_i}}(\gamma)=\frac{z_i^{k_i}\rho_i^{2}\sigma_i^{\beta_i-\frac{1}{2}}\lambda_i^{\varphi_i-\frac{1}{2}} (2\pi)^{1-\frac{\lambda_i+\sigma_i}{2}}}{2(\phi_i\lambda_i)^{k_i}\Gamma(\beta_i)\Gamma(\varphi_i)\gamma}\nonumber \\& G_{\lambda_i+\sigma_i+k_i+1,1+k_i}^{0,\lambda_i+\sigma_i+k_i+1} \bigg[\begin{array}{c}\frac{\lambda_i^{\lambda_i}\sigma_i^{\sigma_i}\Omega_i^{\sigma_i}\Xi_i^{\lambda_i}}{\beta_i^{\sigma_i}\varphi_i^{\lambda_i}} \big(\frac{A_{i}\sqrt{\gamma_{0}}}{\sqrt{\gamma}}\big)^{\phi_i\lambda_i}\end{array} \bigg\vert \begin{array}{c} U_i \\ V_{i} \end{array}\bigg]\nonumber\\&
		\label{eq:snr_pdf_fpt}
	\end{eqnarray}
	where $U_i= \{\mu_i, \{1-\frac{z_i}{\phi_i\lambda_i}\}_{1}^{k_i}\}$ and $V_i= \{\nu_i, \{-\frac{z_i}{\phi_i\lambda_i}\}_{1}^{k_i}\}$, and
	\begin{eqnarray}
		&{F^{\rm FPT}_{\gamma_i}}(\gamma)=\frac{z_i^{k_i}\rho_i^{2}\sigma_i^{\beta_i-\frac{1}{2}}\lambda_i^{\varphi_i-\frac{1}{2}} (2\pi)^{1-\frac{\lambda_i+\sigma_i}{2}}}{2(\phi_i\lambda_i)^{k_i}\Gamma(\beta_i)\Gamma(\varphi_i)}\nonumber \\& H_{\lambda_i+\sigma_i+k_i+2,2+k_i}^{1,\lambda_i+\sigma_i+k_i+1} \bigg[\begin{array}{c}\frac{\lambda_i^{\lambda_i}\sigma_i^{\sigma_i}\Omega_i^{\sigma_i}\Xi_i^{\lambda_i}}{\beta_i^{\sigma_i}\varphi_i^{\lambda_i}} \big(\frac{A_{i}\sqrt{\gamma_{0}}}{\sqrt{\gamma}}\big)^{\phi_i\lambda_i}\end{array} \bigg\vert \begin{array}{c} \tilde{U}_i\\ \tilde{V}_i \end{array}\bigg]\nonumber\\&
		\label{eq:snr_cdf_fpt}
	\end{eqnarray}
	where $\tilde{U}_i= \{(\mu_i,1), (\{1-\frac{z_i}{\phi_i\lambda_i}\}_{1}^{k_i},1), (1,\frac{\phi_i\lambda_i}{2})\}$ and $\tilde{V}_i= \{(0,\frac{\phi_i\lambda_i}{2}),(\nu_i,1), (\{-\frac{z_i}{\phi_i\lambda_i}\}_{1}^{k_i},1)\}$.	
\end{my_theorem}
\begin{IEEEproof}			
	The PDF of the  combined FPT channel  $h_i= h_{fi} h_{tpi}$ can be expressed as  \cite{Papoulis2001}
	\begin{eqnarray}
		f_{h_i}(h_i)=\int_{}^{}\frac{1}{|h_{tpi}|}f_{h_{fi}}({h_i|h_{tpi}}) f_{h_{tpi}}(h_{tpi})dh_{tpi}
		\label{pe_5}
	\end{eqnarray}
	Substituting \eqref{eq:fh} and \eqref{eq:pdf_ha_dgg}  in \eqref{pe_5}, we get
	\begin{eqnarray}
		&f_{h_i}(h_i) = \frac{z_i^{k_i}\rho_i^{2}\sigma_i^{\beta_i-\frac{1}{2}}\lambda_i^{\varphi_i-\frac{1}{2}} (2\pi)^{1-\frac{\lambda_i+\sigma_i}{2}}h_i^{z_i-1}}{\Gamma(k_i)\Gamma(\beta_i)\Gamma(\varphi_i)}\nonumber \\&\int_{h_i}^{\infty}h_{tpi}^{-z_i-1}\left[\ln \left(\frac{h_{tpi}}{h_i}\right)\right]^{k_i-1}\nonumber \\&  G_{\lambda_i+\sigma_i+1,1}^{0,\lambda_i+\sigma_i+1} \bigg[\begin{array}{c}\frac{\lambda_i^{\lambda_i}\sigma_i^{\sigma_i}\Omega_i^{\sigma_i}\Xi^{\lambda_i}}{\beta_i^{\sigma_i}\varphi_i^{\lambda_i}} \big(\frac{A_{i}}{h_{tpi}}\big)^{\phi_i\lambda_i}\end{array} \bigg\vert \begin{array}{c} \mu_i \\ \nu_i \end{array}\bigg] dh_{tpi}\nonumber \\&
		\label{eq_int}
	\end{eqnarray}
	Using the definition of Meijer-G function and interchanging the order of integration, we represent \eqref{eq_int} as
	\begin{eqnarray}
		&f_{h_i}(h_i) = \frac{z_i^{k_i}\rho_i^{2}\sigma_i^{\beta_i-\frac{1}{2}}\lambda_i^{\varphi_i-\frac{1}{2}} (2\pi)^{1-\frac{\lambda_i+\sigma_i}{2}}h_i^{z_i+1}}{\Gamma(k_i)\Gamma(\beta_i)\Gamma(\varphi_i)}\nonumber \\&\frac{1}{2\pi i}\int_{L}^{}\frac{\prod_{j=1}^{\lambda_i+\sigma_i+1}\Gamma(1-\mu_{i,j}+s)}{\Gamma(1-\nu_i+s)}\left(\frac{\lambda_i^{\lambda_i}\sigma_i^{\sigma_i}\Omega_i^{\sigma_i}\Xi_i^{\lambda_i}A^{\phi_i\lambda_i}_i}{\beta_i^{\sigma_i}\varphi_i^{\lambda_i}}\right)^{s}ds\nonumber\\&\int_{h_i}^{\infty}h_{tpi}^{-z_i-s\phi_i\lambda_i-1}\left[\ln \left(\frac{h_{tpi}}{h_i}\right)\right]^{k_i-1} dh_{tpi}
		\label{eq_int1}
	\end{eqnarray}
	where $\mu_{i,j}=1-\frac{\rho_i^2}{\phi_i\lambda_i},\Delta(\sigma_i:1-\beta_i),\Delta(\lambda_i:1-\varphi_i)$. Substituting $\ln(h_{tpi}/h_i)=t$, we solve  the  inner integral of \eqref{eq_int1} in terms of Gamma function:
	\begin{eqnarray}
		&	I_1=\int_{0}^{\infty}t^{k_i-1}e^{-(z_i+s\phi_i\lambda_i)t}dt=\frac{\Gamma(k_i)}{(z_i+s\phi_i\lambda_i)^{k_i}}\nonumber\\&=\frac{\Gamma(k_i)}{(\phi_i\lambda_i)^{k_i}}\left(\frac{\Gamma\left(\frac{z_i}{\phi_i\lambda_i}+s\right)}{\Gamma\left(1+\frac{z_i}{\phi_i\lambda_i}+s\right)}\right)^{k_i}
		\label{I_inner}
	\end{eqnarray}
	Using \eqref{I_inner} in \eqref{eq_int1} and applying the definition of Fox-H function with a transformation of random variable $\gamma_i=\gamma_0h_i^2$,  we get the PDF of SNR in \eqref{eq:snr_pdf_fpt}.
	
	To find the CDF of the SNR   for the FPT channel,  we use 	\eqref{eq:snr_pdf_fpt} in $F^{\rm FPT}_{\gamma_i}(\gamma)=\int_{0}^{\gamma}f^{\rm FPT}_{\gamma_i}(x)dx$, apply the definition of Meijer's G function with the inner integral  $\int_{0}^\gamma \gamma^{-\frac{\phi_i\lambda_is}{2}-1}d\gamma=\frac{\gamma^{-\frac{\phi_i\lambda_is}{2}}}{-\frac{\phi_i\lambda_is}{2}}=\frac{\gamma^{-\frac{\phi_i\lambda_is}{2}}\Gamma\left(-\frac{\phi_i\lambda_is}{2}\right)}{\Gamma\left(1-\frac{\phi_i\lambda_is}{2}\right)}$ to get  \eqref{eq:snr_cdf_fpt}, which concludes the proof of Theorem 1.
\end{IEEEproof}

For shorter OWC links, the effect of atmospheric turbulence can be neglected. Thus, defining $\gamma_i^{\rm FP}=\gamma_0 |h_{fi} h_{pi}|^2$, the PDF of SNR for the FP channel  under the combined effect of random fog and pointing errors for the single OWC link is given as \cite{Esmail2017_Photonics}:
\begin{eqnarray}
&{f^{\rm FP}_{\gamma_i}}(\gamma)= \frac{C^{(1)}}{\sqrt{\gamma\gamma_0}}\left(\sqrt{\frac{\gamma}{\gamma_0}}\right)^{\rho^2_i-1}-\frac{C^{(2)}}{\sqrt{\gamma\gamma_0}}\left(\sqrt{\frac{\gamma}{\gamma_0}}\right)^{\rho^2_i-1}\times\nonumber \\&\Gamma(k_i,m_i \ln(A_i/\sqrt{\gamma/\gamma_0})),~~\gamma\leq A_i^2 \gamma_0
\label{eq:snr_pdf_1}
\end{eqnarray}
where  $C^{(1)}=\frac{z_i^{k_i} \rho^2_i}{2 m_i^{k_i} A_i^{\rho^2_i}}$, $C^{(2)}=\frac{C^{(1)}}{\Gamma(k_i)}$, and $m_i=z_i-\rho^2_i$. 
\begin{my_proposition}
	\label{prop_CDF}
	The CDF of the SNR for the FP channel under the effect of random  fog with pointing errors is given as
	\begin{eqnarray}
	&{F^{\rm FP}_{\gamma_i}}(\gamma)=D^{(1)}\left(\frac{A_i}{\sqrt{\gamma/\gamma_0}}\right)^{-\rho^2_i}-D^{(2)}(k_i-1)!\nonumber\\&\times\sum_{n=0}^{k_i-1}\frac{m_i^n z^{-n-1}_i}{n!}\Gamma \left(n+1,z_i \ln \left(A_i/\sqrt{\gamma/\gamma_0}\right)\right)
	\label{eq:snr_cdf_3}
	\end{eqnarray}                                                                     
	where $D^{(1)}=\frac{z_i^{k_i}}{m_i^{k_i}}$ and $D^{(2)}=\frac{D^{(1)}\rho^2_i}{\Gamma(k_i)}$ are constants. 
\end{my_proposition}

\begin{IEEEproof}
	Substituting $u=\ln \left(\frac{A_i}{\sqrt{\gamma/\gamma_0}}\right)$ in the second term of \eqref{eq:snr_pdf_1}, the CDF ${F^{\rm FP}_{\gamma_i}}(\gamma)= \int_{0}^\gamma f^{\rm FP}_{{\gamma_i}}(\gamma) d \gamma$ is given as:
	\begin{eqnarray}
	{F^{\rm FP}_{\gamma_i}}(\gamma)=D^{(1)}\left(\frac{A_i}{\sqrt{\gamma/\gamma_0}}\right)^{-\rho^2_i}- D^{(2)}\int\limits_{u}^{\infty}e^{-\rho^2_i u}\Gamma[k_i,m_iu]du
	\label{eq:snr_cdf_2}
	\end{eqnarray} 
	Using $\Gamma(a,t) \triangleq (a-1)! e^{-t} \sum_{m=0}^{a-1} \frac{t^m}{m!}$ in \eqref{eq:snr_cdf_2}, and applying the definition of incomplete Gamma function, we get \eqref{eq:snr_cdf_3}.
	
\end{IEEEproof}
It should be noted that the derived CDF for the FP channel in \eqref{eq:snr_cdf_3} is different from  \cite{Esmail2017_Photonics}: it consists of a single incomplete gamma function without the exponential integral, and can useful for performance analysis in  a closed-form using integer $k$.  It can be seen that distribution functions in  \eqref{eq:snr_pdf_1} and \eqref{eq:snr_cdf_3} involves incomplete gamma functions with logarithmic argument requiring novel approaches to performance analysis.  The use of simple  approximation of incomplete Gamma function $\Gamma[k,m\ln u] \approx u^{-m} (m\ln u)^{k-1}$ can simplify the analysis, but the derived approximate expressions grossly overestimate/underestimate the exact performance.

Finally, we discuss the distribution functions with DF relaying. We assume equal transmit power at the source and relay (i.e., $P_t = P_r$) to get the instantaneous SNRs of signals received at the relay and receiver as $\gamma_1$ and $\gamma_2$, respectively. Assuming that $\gamma_1$ and $\gamma_2$ are independent for analytical tractability, the expression of end-to-end SNR for the DF relaying  is given as:
\begin{eqnarray}
\gamma_{}=\min(\gamma_1, \gamma_2)
\label{eq:half_harmonic_mean}
\end{eqnarray}
It is true that $k$ and $\beta$ parameters of the random foggy channel will be identical in both the hops. However,  channel realizations at two distinct points separated over several hundred meters might not be the same. The i.i.d. assumption on the foggy channel for the FP channel has been considered in \cite{Esmail2017_Photonics}.

In general, the CDF and PDF of end-to-end SNR for the DF relaying scheme can be given as \cite{Yang2014_relay}:
\begin{eqnarray}
\Psi(\gamma)= \Psi_1(\gamma)+\Psi_2(\gamma)-\Psi_1(\gamma)\Psi_2(\gamma)
\label{eq:aprox_snr_cdf}
\end{eqnarray}
\begin{eqnarray}
\psi(\gamma)= \psi_1(\gamma)+\psi_2(\gamma)-\psi_1(\gamma)\Psi_2(\gamma)-\psi_2(\gamma)\Psi_1(\gamma)
\label{eq:aprox_snr_pdf}
\end{eqnarray}
where $\psi_1(\gamma)$, $\psi_2(\gamma)$ are the PDF of the first link and second link, respectively. Similarly,  $\Psi_1(\gamma)$ and $\Psi_2(\gamma)$ are the CDF of the first link and the second link, respectively.

\subsection{Outage Probability}	\label{subsec:outage_probability}
Outage probability is a performance measure to demonstrate the effect of the fading channel. It is defined as the probability of
failing to reach  an SNR threshold value $\gamma_{\rm th}$, i.e., $P_{\rm out}=P(\gamma < \gamma_{\rm th})=\Psi(\gamma_{\rm th})$. Thus, exact expressions for the outage probability with FPT and FP channels can be obtained by substituting  \eqref{eq:snr_cdf_fpt} and 	\eqref{eq:snr_cdf_3} with $i=1, 2$ in \eqref{eq:aprox_snr_cdf}, respectively. To derive the asymptotic analysis for the FPT,  we  apply \cite{Kilbas} to get  the outage probability at high SNR $\gamma_0\to \infty$ for the $i$-th link:
\begin{eqnarray}
&P^{\rm FPT,\infty}_{\rm out, i} = \frac{z_i^{k_i}\rho_i^{2}\sigma_i^{\beta_i-\frac{1}{2}}\lambda_i^{\varphi_i-\frac{1}{2}} (2\pi)^{1-\frac{\lambda_i+\sigma_i}{2}}}{2(\phi_i\lambda_i)^{k_i}\Gamma(\beta_i)\Gamma(\varphi_i)}\nonumber \\& \sum_{n=1}^{\lambda_i+\sigma_i+k_i+1}\frac{\Gamma\left((1-a_n)\frac{\phi_i\lambda_i}{2s_n}\right)\prod_{j=1,j\neq n}^{\lambda_i+\sigma_i+k_i+1}\Gamma\left(1-a_j+(a_n-1)\frac{s_j}{s_n}\right)}{s_n\Gamma\left(1-(a_n-1)\frac{\phi_i\lambda_i}{2s_n}\right)\prod_{j=2}^{k_i+2}\Gamma\left(1-b_j+(a_n-1)\frac{t_j}{s_n}\right)}\nonumber\\&\Big(\frac{\lambda_i^{\lambda_i}\sigma_i^{\sigma}\Omega_i^{\sigma_i}\Xi_i^{\lambda_i}}
{\beta_i^{\sigma_i}\varphi_i^{\lambda_i}}
 \big(\frac{A_{i}\sqrt{\gamma_{0}}}{\sqrt{\gamma_{\rm{th}}}}\big)^{\phi_i\lambda_i}\Big)^{\frac{a_n-1}{s_n}}
\label{cdf_h_fpt4}
\end{eqnarray}
where $a_n=a_j=\{\mu_i, \{1-\frac{z_i}{\phi_i\lambda_i}\}_{1}^{k_i}, 1\}$, $s_n=s_j=\{1, 1, \frac{\phi_i\lambda_i}{2}\}$, $b_n=b_j=\{0, \nu_i, \{-\frac{z_i}{\phi_i\lambda_i}\}_{1}^{k_i}\}$ and $t_n=t_j=\{\frac{\phi_i\lambda_i}{2}, 1, 1\}$. 

 The diversity of the FPT channel can be obtained using dominant SNR terms of \eqref{cdf_h_fpt4}. Using the parameters of $a_n$ and $s_n$ in $(\sqrt{\gamma_0})^{\phi_i \lambda_i(\frac{a_n-1)}{s_n}}$ with the exponent of $\gamma_0$ from dominant terms in \eqref{cdf_h_fpt4}, the diversity order of the system is $M_{{\rm out},i}^{\rm FPT}=\min\{\frac{z_i}{2}, \frac{\rho_i^2}{2}, \frac{\alpha_i\beta_i}{2}, \frac{\phi_i\varphi_i}{2}\}$.  Using $i=1, 2$, the diversity order for the dual-hop system with outage probability $P_{\rm out}^{\rm FPT}= P_{{\rm out},1}^{{\rm FPT}}+P_{{\rm out},2}^{{\rm FPT}}-P_{{\rm out},2}^{{\rm FPT}}P_{{\rm out},2}^{{\rm FPT}} $  can be expressed as $M_{{\rm out}}^{\rm FPT}= \min\{\frac{\rho_1^2}{2},\frac{\rho_2^2}{2}, \frac{\alpha_1\beta_1}{2},\frac{\phi_1\varphi_1}{2},  \frac{\alpha_2\beta_2}{2}, \frac{\phi_2\varphi_2}{2}, \frac{z_1}{2}, \frac{z_2}{2} \}$. Thus, the diversity order for the DGG channel with pointing errors as derived in \cite{AlQuwaiee2015, Ashrafzadeh2020} is a special case of the diversity order for the FPT channel. This validates our proposed analysis on the outage probability.  Similarly, using the series expansion of incomplete Gamma function $\Gamma(a,t) \triangleq (a-1)! e^{-t} \sum_{n=0}^{a-1} \frac{t^n}{n!}$\cite{Zwillinger2014} in \eqref{eq:snr_cdf_3}, we  express the outage probability of the $i$-th link  for the FP channel:
	\begin{eqnarray}
	&P^{\rm FP}_{{\rm out}, i}= D_i^{(1)}\left(\frac{A_i^2\gamma_0}{\gamma_{\rm th}}\right)^{-\frac{\rho_i^2}{2}}-D_i^{(2)}(k-1)!\left(\frac{A_i^2\gamma_0}{\gamma_{\rm th}}\right)^{-\frac{z_i}{2}}\nonumber\\&\sum_{n=0}^{k-1}\sum_{j=0}^{n}\frac{m_i^n z_i^{j-n-1}\left(\ln\left(\frac{{A_i}\sqrt{\gamma_0}}{\sqrt{\gamma_{th}}}\right)\right)^{j}}{j!}
	\label{eq:outage_exact_1}
	\end{eqnarray}
 Using \eqref{eq:outage_exact_1} and following the similar steps of  \cite{rahman2020cl}, the diversity order for the FP channel can be derived as $M_{{\rm out},i}^{\rm FP}=  \frac{1}{2}\min\{z_i, \rho_i^2\}$. 

The diversity order  various possibilities of mitigating the impact of pointing errors and fog.  The diversity order provides design criteria of appropriately using the beam width and link distance to mitigate the effect of pointing errors and random fog. As such,   the beam-width (associated with pointing errors) and link distance (associated with fog) can be chosen  to circumvent the fading due to the atmospheric turbulence. For example, under certain conditions   $M_{\rm out}^{\rm FPT}=\frac{z_1}{2}=\frac{2.171}{\beta^{\rm{fog}} d_1}$, and thus the effect of atmospheric turbulence and pointing errors is not dominant and  performance depends only on the fog parameter $\beta^{\rm{fog}}$ and source-to-relay distance $d_1$.
\subsection{Average SNR and Ergodic Rate}\label{subsec:average_snr_rate}
The average SNR and ergodic rate performance are important parameters for the design of communication systems.  In general, expressions for the average SNR and ergodic rate for the IM/DD detector type is given as \cite{Aghajanzadeh2011}:
\begin{eqnarray}
	\label{eq:avg_snr_approx}
 &\bar{\gamma}=\int\limits_{0}^{\zeta}  \gamma \psi(\gamma)d\gamma
\\&\bar{\eta} =\int\limits_{0}^{\zeta}  \log_2(1+\frac{e}{2\pi}\gamma) \psi(\gamma)d\gamma
\label{eq:avg_rate_approx}
\end{eqnarray}
where $\zeta$ is the limit of integration.  It can be easily seen that $\zeta \to \infty$ for the FPT channel due to the effect of atmospheric turbulence.  However, for the FP  channel,  $\zeta=\min (A_1^2\gamma_0, A_2^2\gamma_0)$.  Further, considering the shorter symbol duration (in the range of few nanoseconds), the  OWC  channel can be considered   a slow-fading channel \cite{Esmail2017_Access}. We also analyze the ergodic rate performance  to provide an estimate on the throughput of the system.  We assume that relay requires negligible time to relay the data  while computing the ergodic rate.  It should be noted that there is  a vast literature  on the ergodic rate performance  on the slow fading FSO channels (See \cite{Yang2014_relay, multi_hop_turb2015, dual_hop_turb2017}, and references therein).

In what follows, we derive closed-form expressions of the average SNR and ergodic capacity for the considered relay--assisted system for both FPT and FP fading channels.

\begin{my_lemma}[Average SNR for FPT]
	\label{theorem:snr_fpt}
	If $k_i$ and $\beta_i$ are the parameters of the foggy channel,  $A_i$ and $\rho_i$ are the  parameters of pointing errors,  sets $(\alpha_i,\beta_i,\Omega_i)$, $(\phi_i,\varphi_i,\Xi_i)$ are the parameters for DGG atmospheric turbulence, and $z_i=4.343/\beta^{\rm{fog}} d_i$, then an exact expression of  the average SNR for the FPT channel is
	\begin{eqnarray}
	\bar{\gamma}^{\rm FPT}=\bar{\gamma}^{\rm FPT}_1+\bar{\gamma}^{\rm FPT}_2-\bar{\gamma}^{\rm FPT}_{12}-\bar{\gamma}^{\rm FPT}_{21}
	\end{eqnarray}
	where $\bar{\gamma}^{\rm FPT}_1$ and $\bar{\gamma}^{\rm FPT}_2$ are given in \eqref{avg_snr_final_fpt} with $i=1$ and $i=2$ whereas $\bar{\gamma}^{\rm FPT}_{12}$ and $\bar{\gamma}^{\rm FPT}_{21}$ are given in \eqref{avg_snr_12_fin} and \eqref{avg_snr_21_fin}, respectively.
	\begin{figure*}

\begin{align}
\bar{\gamma}^{\rm FPT}_i = -\frac{z_i^{k_i}\rho_i^{2}\sigma_i^{\beta_i-\frac{1}{2}}\lambda_i^{\varphi_i-\frac{1}{2}} (2\pi)^{1-\frac{\lambda_i+\sigma_i}{2}}}{\phi_i\lambda_i\Gamma(\beta_i)\Gamma(\varphi_i)}\frac{\prod_{j=1}^{\lambda_i+\sigma_i+1}\Gamma\left(1+\frac{2}{\phi_i\lambda_i}-\mu_{i,j}\right)}{(2+z_i)^{k_i}\Gamma\left(1+\frac{2}{\phi_i\lambda_i}-\nu_i\right)}\left(\frac{\lambda_i^{\lambda_i}\sigma_i^{\sigma_i}\Omega_i^{\sigma_i}\Xi_i^{\lambda_i}}{\beta_i^{\sigma_i}\varphi_i^{\lambda_i}} \big(A_{i}\sqrt{\gamma_{0}}\big)^{\phi_i\lambda_i}\right)^{\frac{2}{\phi_i\lambda_i}}
\label{avg_snr_final_fpt}
\end{align}
	
	\begin{eqnarray}
		&\bar{\gamma}^{\rm FPT}_{12} =-\frac{2a_{12}b^{\frac{2}{\phi_1\lambda_1}}_1}{\phi_1\lambda_1} H_{\lambda_2+\sigma_2+k_1+k_2+3,\lambda_1+\sigma_1+k_1+k_2+3}^{\lambda_1+\sigma_1+k_1+2,\lambda_2+\sigma_2+k_2+1} \bigg[\begin{array}{c}\frac{b_2}{b^{\frac{\phi_2\lambda_2}{\phi_1\lambda_1}}_1}\end{array} \bigg\vert \begin{array}{c} U_{12} \\ V_{12} \end{array}\bigg]
		\label{avg_snr_12_fin}
	\end{eqnarray}

	\begin{eqnarray}
	&\bar{\gamma}^{\rm FPT}_{21} =-\frac{2a_{21}b^{\frac{2}{\phi_2\lambda_2}}_2}{\phi_2\lambda_2} H_{\lambda_1+\sigma_1+k_1+k_2+3, \lambda_2+\sigma_2+k_1+k_2+3}^{\lambda_2+\sigma_2+k_2+2, \lambda_1+\sigma_1+k_1+1} \bigg[\begin{array}{c}\frac{b_1}{b^{\frac{\phi_1\lambda_1}{\phi_2\lambda_2}}_2}\end{array} \bigg\vert \begin{array}{c} U_{21} \\ V_{21} \end{array}\bigg]
	\label{avg_snr_21_fin}
\end{eqnarray}
where $U_{12}=\{(1-\nu_1+\frac{2}{\phi_1\lambda_1},\frac{\phi_2\lambda_2}{\phi_1\lambda_1}), (1+\{\frac{z_1}{\phi_1\lambda_1}\}_{1}^{k_1}+\frac{2}{\phi_1\lambda_1},\frac{\phi_2\lambda_2}{\phi_1\lambda_1}), (1,\frac{\phi_2\lambda_2}{2})\}$, $V_{12}=\{(0,\frac{\phi_2\lambda_2}{2}), (\nu_2,1), (\{-\frac{z_2}{\phi_2\lambda_2}\}_{1}^{k_2},1)\}$, $U_{21}=\{(1-\nu_2+\frac{2}{\phi_2\lambda_2},\frac{\phi_1\lambda_1}{\phi_2\lambda_2}), (1+\{\frac{z_2}{\phi_2\lambda_2}\}_{1}^{k_2}+\frac{2}{\phi_2\lambda_2},\frac{\phi_1\lambda_1}{\phi_2\lambda_2}), (1,\frac{\phi_1\lambda_1}{2})\}$, $V_{21}=\{(0,\frac{\phi_1\lambda_1}{2}), (\nu_1,1), (\{-\frac{z_1}{\phi_1\lambda_1}\}_{1}^{k_1},1)\}$, $a_{12}=a_{21}=\frac{{z^{k_1}_1}{z^{k_2}_2}(\rho_1\rho_2)^{2}\sigma_1^{\beta_1-\frac{1}{2}}\sigma_2^{\beta_2-\frac{1}{2}}\lambda_1^{\varphi_1+k_1-\frac{1}{2}}\lambda_2^{\varphi_2+k_2-\frac{3}{2}} (2\pi)^{\frac{4-\lambda_1+\sigma_1-\lambda_2+\sigma_2}{2}}\phi_1^{k_1}\phi_2^{k_2-1}}{4\Gamma(\beta_1)\Gamma(\beta_2)\Gamma(\varphi_1)\Gamma(\varphi_2)}$, and $b_{12}=b_{21}=\frac{\lambda_1^{\lambda_1}\lambda_2^{\lambda_2}\sigma_1^{\sigma_1}\sigma_2^{\sigma_2}\Omega_1^{\sigma_1}\Omega_2^{\sigma_2}\Xi_1^{\lambda_1}\Xi_2^{\lambda_2}A^{\phi_1\lambda_1}_1A^{\phi_2\lambda_2}_2\gamma^{\frac{\phi_1\lambda_1+\phi_2\lambda_2}{2}}_0}{\beta_1^{\sigma_1}\beta_2^{\sigma_2}\varphi_1^{\lambda_1}\varphi_2^{\lambda_2}}$.
\hrule

\end{figure*}

	\end{my_lemma}

\begin{IEEEproof}
Using	\eqref{eq:snr_pdf_fpt} in \eqref{eq:avg_snr_approx} with the substitution  $\gamma^{-\frac{\phi_i\lambda_i}{2}}=t$, and applying the identity [\cite{Wolfram}, eq. 07.34.21.0009.01] of the single Meijer-G with few simplifications, we get \eqref{avg_snr_final_fpt}. Next, we use $\psi(\gamma)=f^{\rm FPT}_{\gamma_1}(\gamma)F^{\rm FPT}_{\gamma_2}(\gamma)$ in \eqref{eq:avg_snr_approx} with a substitution $\gamma^{-\frac{\phi_1\lambda_1}{2}}=t$ to get

\begin{eqnarray}
&\bar{\gamma}^{\rm FPT}_{12} =-\frac{2a_{12}}{\phi_1\lambda_1}\int_{0}^{\infty}t^{-\frac{2}{\phi_1\lambda_1}-1}\nonumber \\& G_{\lambda_1+\sigma_1+k_1+1,1+k_1}^{0,\lambda_1+\sigma_1+k_1+1} \bigg[\begin{array}{c}b_1t\end{array} \bigg\vert \begin{array}{c} U_1 \\ V_{1} \end{array}\bigg]\nonumber\\& H_{\lambda_2+\sigma_2+k_2+2,2+k_2}^{1,\lambda_2+\sigma_2+k_2+1} \bigg[\begin{array}{c}b_2t^{\frac{\phi_2\lambda_2}{\phi_1\lambda_1}}\end{array} \bigg\vert \begin{array}{c} \tilde{U}_2\\ \tilde{V}_2 \end{array}\bigg]dt
\label{avg_snr_12_3}
\end{eqnarray}
where $b_1=\frac{z_1^{k_1}\rho_1^{2}\sigma_1^{\beta_1-\frac{1}{2}}\lambda_1^{\varphi_1-\frac{1}{2}} (2\pi)^{1-\frac{\lambda_1+\sigma_1}{2}}}{2(\phi_1\lambda_1)^{k_1}\Gamma(\beta_1)\Gamma(\varphi_1)}$, $b_2=\frac{z_2^{k_2}\rho_2^{2}\sigma_2^{\beta_2-\frac{1}{2}}\lambda_2^{\varphi_2-\frac{1}{2}} (2\pi)^{1-\frac{\lambda_2+\sigma_2}{2}}}{2(\phi_2\lambda_2)^{k_2}\Gamma(\beta_2)\Gamma(\varphi_2)}$, $U_1= \{\mu_1, \{1-\frac{z_1}{\phi_1\lambda_1}\}_{1}^{k_1}\}$, $V_1= \{\nu_1, \{-\frac{z_1}{\phi_1\lambda_1}\}_{1}^{k_1}\}$, $\tilde{U}_2= \{(\mu_2,1), (\{1-\frac{z_2}{\phi_2\lambda_2}\}_{1}^{k_2},1), (1,\frac{\phi_2\lambda_2}{2})\}$ and $\tilde{V}_2= \{(0,\frac{\phi_2\lambda_2}{2}),(\nu_2,1), (\{-\frac{z_2}{\phi_2\lambda_2}\}_{1}^{k_2},1)\}$.

Converting the  Meijer-G function to Fox-H function and applying the identity [\cite{Wolfram}, eq. 07.34.21.0012.01
] in \eqref{avg_snr_12_3}, we get 	\eqref{avg_snr_12_fin}. We apply the similar procedure to derive  $\bar{\gamma}^{\rm FPT}_{21}$ in \eqref{avg_snr_21_fin} with a substitution $\gamma^{-\frac{\phi_2\lambda_2}{2}}=t$, which concludes the proof.
\end{IEEEproof}	
\begin{my_lemma}[Ergodic rate for FPT]
	\label{theorem:rate_fpt}
	If $k_i$ and $\beta_i$ are the parameters of the foggy channel,  $A_i$ $\rho_i$ are the  parameters of pointing errors,  sets $(\alpha_i,\beta_i,\Omega_i)$, $(\phi_i,\varphi_i,\Xi_i)$ are the parameters for atmospheric turbulence, and $z_i=4.343/\beta^{\rm{fog}} d_i$, then an exact expression of  the ergodic rate for the FPT channel is
	\begin{eqnarray}
		\bar{\eta}^{\rm FPT}=\bar{\eta}^{\rm FPT}_1+\bar{\eta}^{\rm FPT}_2-\bar{\eta}^{\rm FPT}_{12}-\bar{\eta}^{\rm FPT}_{21}
	\end{eqnarray}
	where $\bar{\eta}^{\rm FPT}_1$ and $\bar{\eta}^{\rm FPT}_2$ are given in \eqref{avg_rate_final_fpt} with $i=1$ and $i=2$ whereas $\bar{\eta}^{\rm FPT}_{12}$ and $\bar{\eta}^{\rm FPT}_{21}$ are given in \eqref{avg_rate_12_fin} and \eqref{avg_rate_21_fin}, respectively.
	\end{my_lemma}
\begin{IEEEproof}
Using \eqref{eq:snr_pdf_fpt} in \eqref{eq:avg_snr_approx} with $\ln(1+\frac{e}{2\pi}\gamma)=G_{2,2}^{1,2} \bigg[\begin{array}{c}\frac{e}{2\pi}\gamma\end{array} \bigg\vert \begin{array}{c} 1,1 \\ 1,0 \end{array}\bigg]$, we get 
\begin{eqnarray}
&\bar{\eta}_i = \frac{z_i^{k_i}\rho_i^{2}\sigma_i^{\beta_i-\frac{1}{2}}\lambda_i^{\varphi_i-\frac{1}{2}} (2\pi)^{1-\frac{\lambda_i+\sigma_i}{2}}}{\ln(4)(\phi_i\lambda_i)^{k_i}\Gamma(\beta_i)\Gamma(\varphi_i)} \nonumber \\&\int_{0}^{\infty}\gamma^{-1} G_{2,2}^{1,2} \bigg[\begin{array}{c}\frac{e}{2\pi}\gamma\end{array} \bigg\vert \begin{array}{c} 1,1 \\ 1,0 \end{array}\bigg] \nonumber \\& G_{\lambda_i+\sigma_i+k_i+1,1+k_i}^{0,\lambda_i+\sigma_i+k_i+1} \bigg[\begin{array}{c}\frac{\lambda_i^{\lambda_i}\sigma_i^{\sigma_i}\Omega_i^{\sigma_i}\Xi_i^{\lambda_i}}{\beta_i^{\sigma_i}\varphi_i^{\lambda_i}} \big(\frac{A_{i}\sqrt{\gamma_{0}}}{\sqrt{\gamma}}\big)^{\phi_i\lambda_i}\end{array} \bigg\vert \begin{array}{c} U_i \\ V_{i} \end{array}\bigg]d\gamma\nonumber\\&
\label{rate_fpt1}
\end{eqnarray}
We apply  the identity [\cite{Wolfram}, eq. 07.34.21.0012.01
] in \eqref{rate_fpt1} to get \eqref{avg_rate_final_fpt}. 
To derive $\bar{\eta}_{12}$, we use $\psi(\gamma)=f^{\rm FPT}_{\gamma_1}(\gamma)F^{\rm FPT}_{\gamma_2}(\gamma)$ in \eqref{eq:avg_snr_approx} with  Meijer-G equivalent of $\log(1+\frac{e}{2\pi})$, and apply the definition of Meijer-G and Fox-H functions \cite{Kilbas} to get
\begin{eqnarray}
&\bar{\eta}_{12} = \frac{a_{12}}{\log(16)}\frac{1}{2\pi i}\int_{L}^{}\frac{\Gamma\left(\frac{z_1}{\phi_1\lambda_1}+s\right)\prod_{j=1}^{\lambda_1+\sigma_1+1}\Gamma(1+\mu_{1,j}+s)}{\Gamma(1+\nu_1+s)\Gamma\left(1+\frac{z_1}{\phi_1\lambda_1}+s\right)}\nonumber\\&\frac{\Gamma\left(\frac{z_2}{\phi_2\lambda_2}+s\right)\Gamma(-s\frac{\phi_2\lambda_2}{2})\prod_{j=1}^{\lambda_2+\sigma_2+1}\Gamma(1+\mu_{2,j}+s)}{\Gamma(1+\nu_2+s)\Gamma(1-s\frac{\phi_2\lambda_2}{2})\Gamma\left(1+\frac{z_2}{\phi_2\lambda_2}+s\right)}b_{12}^sds\nonumber\\&\int_{0}^{\infty}\gamma^{-1-s\left(\frac{\phi_1\lambda_1+\phi_2\lambda_2}{2}\right)}\log\left(1+\frac{e}{2\pi}\gamma\right)d\gamma
\label{rate_12_1}
\end{eqnarray}

Solving the inner integral
\begin{eqnarray}\label{eq:inner:eta12}
&I=\int_{0}^{\infty}\gamma^{-1-s\left(\frac{\phi_1\lambda_1+\phi_2\lambda_2}{2}\right)}\log\left(1+\frac{e}{2\pi}\gamma\right)d\gamma\nonumber\\&=\frac{\left(\frac{e}{2\pi}\right)^{s\left(\frac{\phi_1\lambda_1+\phi_2\lambda_2}{2}\right)}\pi \text{Csc}[s\left(\frac{\phi_1\lambda_1+\phi_2\lambda_2}{2}\right)\pi]}{s\left(\frac{\phi_1\lambda_1+\phi_2\lambda_2}{2}\right)}\nonumber\\&=\frac{\left(\frac{e}{2\pi}\right)^{s\left(\frac{\phi_1\lambda_1+\phi_2\lambda_2}{2}\right)}}{s\left(\frac{\phi_1\lambda_1+\phi_2\lambda_2}{2}\right)}\Gamma\left(s\left(\frac{\phi_1\lambda_1+\phi_2\lambda_2}{2}\right)\right)\nonumber\\&\Gamma\left(1-s\left(\frac{\phi_1\lambda_1+\phi_2\lambda_2}{2}\right)\right)=\left(\frac{e}{2\pi}\right)^{s\left(\frac{\phi_1\lambda_1+\phi_2\lambda_2}{2}\right)}\nonumber\\&\frac{\Gamma(s\left(\frac{\phi_1\lambda_1+\phi_2\lambda_2}{2}\right))\Gamma(s\left(\frac{\phi_1\lambda_1+\phi_2\lambda_2}{2}\right))\Gamma\left(1-s\left(\frac{\phi_1\lambda_1+\phi_2\lambda_2}{2}\right)\right)}{\Gamma\left(1+s\left(\frac{\phi_1\lambda_1+\phi_2\lambda_2}{2}\right)\right)}
\end{eqnarray}
and using \eqref{eq:inner:eta12} in \eqref{rate_12_1} with the application of  the definition of Fox-H function, we get \eqref{avg_rate_12_fin}.  We apply a similar procedure to derive  $\bar{\eta}^{\rm FPT}_{21}$ in \eqref{avg_rate_21_fin}, which concludes the proof of Lemma.

\end{IEEEproof}
	\begin{figure*}
	\begin{align}
	\bar{\eta}^{\rm FPT}_i = \frac{z_i^{k_i}\rho_i^{2}\sigma_i^{\beta_i-\frac{1}{2}}\lambda_i^{\varphi_i-\frac{1}{2}} (2\pi)^{1-\frac{\lambda_i+\sigma_i}{2}}}{\ln(4)(\phi_i\lambda_i)^{k_i}\Gamma(\beta_i)\Gamma(\varphi_i)}H_{\lambda_i+\sigma_i+k_i+3,3+k_i}^{2,\lambda_i+\sigma_i+k_i+2} \bigg[\begin{array}{c}b_i\left(\frac{e}{2\pi}\right)^{\frac{\phi_i\lambda_i}{2}}\end{array} \bigg\vert \begin{array}{c} \hat{U} \\ \hat{V} \end{array}\bigg]
	\label{avg_rate_final_fpt}
	\end{align}
	where $\hat{b}_i=\frac{\lambda_i^{\lambda_i}\sigma_i^{\sigma_i}\Omega_i^{\sigma_i}\Xi_i^{\lambda_i}}{\beta_i^{\sigma_i}\varphi_i^{\lambda_i}} \big(A_{i}\sqrt{\gamma_{0}}\big)^{\phi_i\lambda_i}$, $\hat{U}=(\mu_i, 1), (\{1-\frac{z_i}{\phi_i\lambda_i}\}_{1}^{k_i},1), (0,-\frac{\phi_i\lambda_i}{2}), (1,-\frac{\phi_i\lambda_i}{2})$, $\hat{V}=(\nu_i,1),(\{-\frac{z_i}{\phi_i\lambda_i}\}_{1}^{k_i},1)$.	
	\begin{eqnarray}
	&\bar{\eta}^{\rm FPT}_{12} =\frac{\hat{a}_{12}}{\log(2)}H_{\lambda_1+\sigma_1+k_1+\lambda_2+\sigma_2+k_2+4,k_1+k_2+4}^{1,\lambda_1+\sigma_1+k_1+\lambda_2+\sigma_2+k_2+4} \bigg[\begin{array}{c}\hat{b}_{12}\left(\frac{e}{2\pi}\right)^{\frac{\phi_1\lambda_1+\phi_2\lambda_2}{2}}\end{array} \bigg\vert \begin{array}{c} \hat{U}_{12} \\ \hat{V}_{12} \end{array}\bigg]
	\label{avg_rate_12_fin}
	\end{eqnarray}
	
	\begin{eqnarray}
	&\bar{\eta}^{\rm FPT}_{21} =\frac{\hat{a}_{21}}{\log(2)}H_{\lambda_1+\sigma_1+k_1+\lambda_2+\sigma_2+k_2+4,k_1+k_2+4}^{1,\lambda_1+\sigma_1+k_1+\lambda_2+\sigma_2+k_2+4} \bigg[\begin{array}{c}\hat{b}_{21}\left(\frac{e}{2\pi}\right)^{\frac{\phi_1\lambda_1+\phi_2\lambda_2}{2}}\end{array} \bigg\vert \begin{array}{c} \hat{U}_{21} \\ \hat{V}_{21} \end{array}\bigg]
	\label{avg_rate_21_fin}
	\end{eqnarray}
	where $\hat{U}_{12}=\{(\mu_1,1),(\mu_2,1), (\{1-\frac{z_1}{\phi_1\lambda_1}\}_{1}^{k_1},1),(\{1-\frac{z_2}{\phi_2\lambda_2}\}_{1}^{k_2},1),(0,\frac{\phi_1\lambda_1+\phi_2\lambda_2}{2}), (0,\frac{\phi_1\lambda_1+\phi_2\lambda_2}{2}), (1, \frac{\phi_2\lambda_2}{2})\}$, $\hat{V}_{12}=\{(0, \frac{\phi_2\lambda_2}{2}), (1,\frac{\phi_1\lambda_1+\phi_2\lambda_2}{2}), (\nu_1,1),(\nu_2,1), (\{-\frac{z_1}{\phi_1\lambda_1}\}_{1}^{k_1},1),(\{-\frac{z_2}{\phi_2\lambda_2}\}_{1}^{k_2},1), (0,\frac{\phi_1\lambda_1+\phi_2\lambda_2}{2})\}$, $\hat{U}_{21}=\{(\mu_1,1),(\mu_2,1), (\{1-\frac{z_1}{\phi_1\lambda_1}\}_{1}^{k_1},1),(\{1-\frac{z_2}{\phi_2\lambda_2}\}_{1}^{k_2},1),(0,\frac{\phi_1\lambda_1+\phi_2\lambda_2}{2}), (0,\frac{\phi_1\lambda_1+\phi_2\lambda_2}{2}), (1, \frac{\phi_1\lambda_1}{2})\}$, $\hat{V}_{21}=\{(0, \frac{\phi_1\lambda_1}{2}), (1,\frac{\phi_1\lambda_1+\phi_2\lambda_2}{2}), (\nu_1,1),(\nu_2,1), (\{-\frac{z_1}{\phi_1\lambda_1}\}_{1}^{k_1},1),(\{-\frac{z_2}{\phi_2\lambda_2}\}_{1}^{k_2},1), (0,\frac{\phi_1\lambda_1+\phi_2\lambda_2}{2})\}$,  $\hat{a}_{12}=\hat{a}_{21}=\frac{{z^{k_1}_1}{z^{k_2}_2}(\rho_1\rho_2)^{2}\sigma_1^{\beta_1-\frac{1}{2}}\sigma_2^{\beta_2-\frac{1}{2}}\lambda_1^{\varphi_1+k_1-\frac{1}{2}}\lambda_2^{\varphi_2+k_2-\frac{3}{2}} (2\pi)^{\frac{4-\lambda_1+\sigma_1-\lambda_2+\sigma_2}{2}}\phi_1^{k_1}\phi_2^{k_2-1}}{4\Gamma(\beta_1)\Gamma(\beta_2)\Gamma(\varphi_1)\Gamma(\varphi_2)}$, and $\hat{b}_{12}=\hat{b}_{21}=\frac{\lambda_1^{\lambda_1}\lambda_2^{\lambda_2}\sigma_1^{\sigma_1}\sigma_2^{\sigma_2}\Omega_1^{\sigma_1}\Omega_2^{\sigma_2}\Xi_1^{\lambda_1}\Xi_2^{\lambda_2}A^{\phi_1\lambda_1}_1A^{\phi_2\lambda_2}_2\gamma^{\frac{\phi_1\lambda_1+\phi_2\lambda_2}{2}}_0}{\beta_1^{\sigma_1}\beta_2^{\sigma_2}\varphi_1^{\lambda_1}\varphi_2^{\lambda_2}}$.
	\hrule

\end{figure*}

As explained earlier, the atmospheric turbulence can be neglected for shorter OWC links. In the following, we derive closed-form expressions of the average SNR and ergodic for the FP channels by considering	the i.i.d. fading of the random fog channel. In this scenario, the pointing errors are assumed to be i.ni.d for both the links.
\begin{my_lemma}[Average SNR and ergodic rate for i.ni.d  FP]
	\label{theorem:snr_fp}
	If $k$ and $\beta^{\rm{fog}}$ are the parameters of the foggy channel,  $A_1$, $A_2$, $\rho_1$, and $\rho_2$ are the channel parameters of pointing errors, and $z_1=4.343/\beta^{\rm{fog}} d_1$, $z_2=4.343/\beta^{\rm{fog}} d_2$ with $d_2\geq d_1$, then a closed-form expression of the average SNR and a lower bound on ergodic rate are given by
	\begin{eqnarray}
	&\bar{\gamma}^{\rm FP}={\cal{F}}_{\gamma}(A_1,A_2,\rho_1,\rho_2,z_1,z_2,k)
\\&\bar{\eta}^{\rm FP}={\cal{F}}_{\eta}(A_1,A_2,\rho_1,\rho_2,z_1,z_2,k)
	\end{eqnarray}
	\end{my_lemma}
\begin{IEEEproof}	
Assuming longer second link, we use the upper limit of the integration as $A_2^2 \gamma_0$ since $d_2\geq d_1 \rightarrow A_2\leq A_1$.  We substitute \eqref{eq:snr_pdf_1} and \eqref{eq:snr_cdf_3} in  \eqref{eq:aprox_snr_pdf} to get $f(\gamma)$ in  terms of system parameters. Since each $f_1(\gamma)$,  $f_2(\gamma)$,  $F_1(\gamma)$, and $F_2(\gamma)$ consist of two terms resulting into twelve terms of integration in  \eqref{eq:avg_snr_approx}.

 We use $m_1=(z_1-\rho_1^2)$, $m_2=(z_2-\rho_2^2)$ and  the series expansion $\Gamma(a,t) \triangleq (a-1)! e^{-t} \sum_{m=0}^{a-1} \frac{t^m}{m!}$, and substitute $u=\frac{A_1}{\sqrt{\gamma/\gamma_0}}$, and $v=\frac{A_2}{\sqrt{\gamma/\gamma_0}}$ to simplify the integrations into  algebraic functions. Apart from simple integrations, we also encounter the following integration terms:
\begin{align}
\begin{split}
\small
\int\limits_{1}^{\infty} u^{-n}\left(\ln(u)\right)^{p} du, \hspace{-1mm} \int\limits_{a}^{\infty} u^{-n-3}\left(\ln(u)\right)^{p} du, \hspace{-1mm} \int\limits_{1}^{\infty} u^{-n} \Gamma[k,n\ln u] du
\label{eq:identity_1_asmyt2}
\end{split}
\end{align}
We solve the integration of  \eqref{eq:identity_1_asmyt2} in closed-forms, as represented in  Appendix B.  Using \eqref{eq:identity_1_asmyt}, \eqref{eq:identity_2_asmyt} and \eqref{eq:identity_3_asmyt} of Appendix B in \eqref{eq:avg_snr_approx} with some algebraic simplifications, we can get the average SNR in a closed-form (not presented due to the space constraint). Similarly, to obtain the ergodic rate, we use the inequality $\log_2(1+\gamma)\geq\log_2(\gamma)$  in \eqref{eq:avg_rate_approx} and follow the same procedure used in deriving the average SNR expression. Here, in addition to terms in \eqref{eq:identity_1_asmyt2}, we also need to integrate the following:
\begin{align}
\begin{split}
\int\limits_{1}^{\infty} u^{-n}\left(\ln(au)\right)^{p} \left(\ln(u)\right)^{t} du
\label{eq:identity_4_asmyt2}
\end{split}
\end{align}
A closed-form expression of \eqref{eq:identity_4_asmyt2} is given in Appendix A. Thus, using \eqref{eq:identity_1_asmyt}, \eqref{eq:identity_2_asmyt},  \eqref{eq:identity_3_asmyt}, and  \eqref{eq:identity_4_asmyt} of Appendix A in \eqref{eq:avg_rate_approx}, we get can get a closed-form expression of the ergodic rate (not presented due to the space constraint).
\end{IEEEproof}	
 In the following Lemma 1,  we present the closed-form expressions by  simplifying the derived expression by considering an i.i.d  fading model  and when the  relay is located in the middle of the source and the destination. 
\begin{my_lemma}[Average SNR and ergodic rate for i.i.d  FP]
	\label{lemma:snr_symt}
	If $k$ and $\beta^{\rm{fog}}$ are  the parameters of the foggy channel,  $A$ and $\rho$ are the parameters of pointing errors, and a relay is at the mid-point with $z=8.686/\beta^{\rm{fog}} d$,  then a  closed-form expression of of average SNR and a lower bound on the ergodic rate are:
		\begin{eqnarray}\label{func:symt_snr}
&\bar{\gamma}^{\rm FP}={\cal{F}}_{\gamma}(A,\rho,z,k)\\&\bar{\eta}^{\rm FP}={\cal{F}}_{\eta}(A,\rho,z,k)
\label{func:symt_rate}
\end{eqnarray}	
	where ${\cal{F}}_{\gamma}(A,\rho,z,k)$ and ${\cal{F}}_{\eta}(A,\rho,z,k)$ are given in \eqref{eq:close_form_snr_symt_1} and \eqref{eq:close_form_rate_symt_1} respectively.			
	\end{my_lemma}

\begin{IEEEproof}
Using \eqref{eq:aprox_snr_pdf} in  \eqref{eq:avg_snr_approx} and \eqref{eq:avg_rate_approx}, and noting that $A_1=A_2=A$, and  $\rho_1=\rho_2=\rho$ under the symmetric i.i.d fading model,  we get:
\begin{eqnarray}
\label{eq:snr_smyt_proof}
&\bar{\gamma}= 2 \int\limits_{0}^{A^2\gamma_0}  \gamma \left[f_\gamma(\gamma)-f_\gamma(\gamma)F_\gamma(\gamma) \right]d\gamma\\&\bar{\eta}= 2 \int\limits_{0}^{A^2\gamma_0}  \log_2(1+\frac{e}{2\pi}\gamma) \left[f_\gamma(\gamma)-f_\gamma(\gamma)F_\gamma(\gamma) \right]d\gamma
\label{eq:rate_smyt_proof}
\end{eqnarray}		
Substituting $u=\frac{A}{\sqrt{{\gamma}/{\gamma_0}}}$, $m=(z_r-\rho^2)$ and using series expansion $\Gamma(a,t) \triangleq (a-1)! e^{-t} \sum_{m=0}^{a-1} \frac{t^m}{m!}$ in \eqref{eq:snr_smyt_proof} for the average SNR and \eqref{eq:rate_smyt_proof} with the inequality $\log_2(1+\gamma)\geq\log_2(\gamma)$ for the ergodic rate, we encounter some simple integration terms along with the first and third integration terms of \eqref{eq:identity_1_asmyt2}.  Using \eqref{eq:identity_1_asmyt} and \eqref{eq:identity_3_asmyt} of Appendix B  with some algebraic simplifications, we get \eqref{func:symt_snr} and  \eqref{func:symt_rate} of Lemma \ref{lemma:snr_symt}. 
\end{IEEEproof} 
\begin{figure*}[tp]
	\begin{eqnarray}
	&{\cal{F}}_{\gamma}(A,\rho,z,k)= 2\bigg(\frac{2C^{(1)}A^{2+\rho^2}\gamma_0}{(2+\rho^2)}+2C^{(2)}A^{2+\rho^2}\gamma_0\frac{\left(-1+\left(\frac{z-\rho^2}{2+z}\right)^k\right)\Gamma[k]}{(2+\rho^2)}-\frac{C^{(1)}D^{(1)}A^{2+\phi^2}\gamma_0}{(1+\rho^2)}+C^{(1)}D^{(2)}(k-1)!\nonumber\\&\times\sum_{i=0}^{k-1}\frac{m^iz^{-i-1}}{i!}\frac{\Gamma (i+1) \left(\rho^2+z+2\right)^{-i-1} \left(\left(\rho^2+z+2\right)^{i+1}-z^{i+1}\right)}{\rho^2+2}-C^{(2)}D^{(1)}A^{2+\rho^2}\gamma_0\frac{\left(-1+\left(\frac{m}{2+m+2\rho^2}\right)^k\right)\Gamma[k]}{(1+\rho^2)}\nonumber\\&-C^{(2)}D^{(2)}A^{\rho^2+2}i!((k-1)!)^2\sum_{i=0}^{k-1}\sum_{j=0}^{k-1}\sum_{t=0}^{i}\frac{m^{i+j}z^{t-i-1}\Gamma (j+t+1)}{i!j!t!\left(\rho^2+2\right)^{j+t+1}} \bigg)
	\label{eq:close_form_snr_symt_1}
	\end{eqnarray}
\end{figure*}

\begin{figure*}[tp]
	\begin{eqnarray}
	&{\cal{F}}_{\eta}(A,\rho,z,k)= \log_2(\frac{e}{2\pi })+2.8854\bigg(\frac{2C^{(1)}A^{\rho^2}\left(-2+\rho^2\ln (A^2\gamma_0)\right)}{\rho^4}-2C^{(2)}A^{\phi^2}(k-1)!\sum_{i=0}^{k-1}\frac{m^i\left(-2(1+i)+z\ln (A^2\gamma_0)\right)}{z^{2+i}}-\nonumber\\&\frac{C^{(1)}D^{(1)}A^{\rho^2}\left(-1+\rho^2\ln (A^2\gamma_0)\right)}{\rho^4}+2C^{(1)}D^{(2)}A^{\rho^2}(k-1)!\sum_{i=0}^{k-1}\sum_{j=0}^{i}\frac{m^iz^{j-i-1}\left(-2(1+j)+(\rho^2+z)\ln (A^2\gamma_0)\right)}{(\rho^2+z)^{2+j}}+\nonumber\\&2C^{(2)}D^{(1)}A^{\rho^2}(k-1)!\sum_{i=0}^{k-1}\frac{m^i\left(-2(1+i)+(\rho^2+z)\ln (A^2\gamma_0)\right)}{(\rho^2+z)^{2+i}}+\nonumber\\&C^{(2)}D^{(2)}A^{\rho^2}((k-1)!)^2\sum_{i=0}^{k-1}\sum_{j=0}^{k-1}\sum_{t=0}^{i}\frac{m^{i+j}z^{-i-j-3}(j+t)!\left(1+j+t-z\ln (A^2\gamma_0)\right)}{j!t!2^{j+t}}\bigg)	
	\label{eq:close_form_rate_symt_1}
	\end{eqnarray}
	\hrulefill
\end{figure*}

Further, we consider light foggy conditions  (i.e., $k=2$) to derive  simple analytical expressions on the average SNR and ergodic rate for the i.i.d model in the following Corollary.
\begin{my_corollary}[Average SNR and ergodic rate for i.i.d  FP with $k=2$]
	\label{corollary_snr_rate_exact_k2}
	If $k=2$ and $\beta^{\rm{fog}}$ are  the parameters of the foggy channel,  $A$ and $\rho$ are the parameters of pointing errors, and relay is at the mid-point with $z=8.686/\beta^{\rm{fog}} d$,  then a  closed-form expression of average SNR and a lower bound on the ergodic rate are given as
	\begin{eqnarray}
		&\bar{\gamma}^{\rm FP}= 2 (A \rho z)^2\gamma_0\Big(\frac{1}{(2+\rho^2)(2+z)^2}-\nonumber\\&\frac{(2 (1 + z)^3 + \rho^4 (1 + 2 z) + \rho^2 (3 + 4 z (2 + z)))}{4 (1 + \rho^2) (1 + z)^3 (2 + \rho^2 + z)^2}\Big)
	\label{eq:snr_k2}
	\end{eqnarray}
	\begin{eqnarray}
	&\bar{\eta}^{\rm FP} \geq  \log_2(\frac{e}{2\pi })+ 2\Big(\big(\frac{\rho^2 z (2 \ln a + \ln \gamma_0)-2 (2 \rho^2 + z)}{\rho^2 z \ln 2}\big)\nonumber \\ &-0.36\big(\rho^2(\rho^2 + z)^{-2}-2\rho^{-2} - 5z^{-1}  - \nonumber \\&3(\rho^2 + z)^{-1}+ 4 \ln A + 2 \ln \gamma_0\big)\Big)
	\label{eq:rate_k2}
	\end{eqnarray}
\end{my_corollary}
\begin{IEEEproof}
The proof follows the similar procedure used in Lemma \ref{lemma:snr_symt} with $k=2$.
\end{IEEEproof}

In our earlier work \cite{rahman2020cl}, we have shown that the average SNR without relaying is  $\bar{\gamma}_{\rm direct}= \frac{(z A \rho)^2 \gamma_0}{(2+\rho^2)(2+z_{\rm direct})^2}$, where $z_{\rm direct}=4.343/\beta^{\rm{fog}} d $  and $d$ is the link distance between the source and destination. Thus,  the first term in \eqref{eq:snr_k2}  corresponds to twice of  the average SNR  without relaying.  Since the negative term in \eqref{eq:snr_k2} is insignificant, we expect a higher average SNR with relaying. Similar conclusions can be made for the ergodic rate performance. These have been extensively studied through numerical analysis in Section IV.

Finally, we demonstrate the impact of randomness in the path gain due to the fog by considering that the fog causes a deterministic path gain $L = e^{-\tau d_{}}$, where $d_{}$ is the link distance (in \mbox{km}) and $\tau$ is the atmospheric attenuation factor, which depends  on the visibility range and may depend on the wavelength \cite{Kim2001}. 
\begin{my_corollary}[Average SNR and ergodic rate for i.i.d  FP channels with deterministic path gain]
	\label{proposition_snr_rate_asymptotic}
	If the fog  causes a deterministic path gain  $L_r= e^{-\tau d/2} $ with relay-assisted transmission  $d_1=d_2= d/2$ for an i.i.d fading model  under random pointing errors,  expressions of average SNR and ergodic rate are given as
		\begin{align}
			\begin{split}
		\bar{\gamma}^{\rm FP}= \frac{(AL_r)^2\rho^4\gamma_0}{(2+3\rho^2+\rho^4)},\text{and }\\~\bar{\eta}^{\rm FP}=  \log_2 \left(\frac{e}{2\pi }\right)+\frac{1.4427(\rho^2\ln((AL_r)^2\gamma_0)-3)}{\rho^2}
		\end{split}
	\label{eq:snr_rate_asymp}
	\end{align}
\end{my_corollary}
\begin{IEEEproof}
Substituting $h_{pt}=\sqrt{\frac{\gamma}{\gamma_0}}$ in \eqref{eq:pdf_hp}, we get an asymptotic  PDF of the SNR for an OWC system under the combined effect of  atmospheric turbulence and pointing errors with atmospheric path gain:
	\begin{eqnarray}
	f_{\gamma}(\gamma)= \frac{\rho^2}{2\sqrt{ \gamma \gamma_{0}} (AL_r)^{\rho^2} } \left(\sqrt{\frac{\gamma}{\gamma_{0}}}\right) ^{\rho^2-1},  0\leq\gamma\leq (AL_r)^2\gamma_0
	\label{eq:pdf_SNR_atm_point_asymp}
	\end{eqnarray}
Using \eqref{eq:pdf_SNR_atm_point_asymp} and the CDF $F_\gamma(\gamma)=\int_{0}^{\gamma}f(\gamma)d\gamma$  in \eqref{eq:snr_smyt_proof} and \eqref{eq:rate_smyt_proof}, it is straightforward to prove \eqref{eq:snr_rate_asymp}.
\end{IEEEproof}

Comparing 	\eqref{eq:close_form_snr_symt_1} and 	\eqref{eq:close_form_rate_symt_1} with the  expressions in 	\eqref{eq:snr_rate_asymp} shows that the randomness in fog significantly complicates  the system analysis. Further, the attenuation coefficient using  deterministic path gain may overestimate/underestimate the performance obtained  with the random fog distribution.

	\subsection{Average BER}

In this subsection, we derive  the average BER for the proposed relay-assisted scheme. Assuming IM/DD, an unified expressions of  the average BER  is given as \cite{dual_hop_turb2017}:
	\begin{eqnarray} \label{eq:ber}
		\bar{P}_{e} = \frac{\delta}{2\Gamma(p)}\sum_{n=1}^{N}q_n^p\int_{0}^{\infty} \gamma^{p-1} {e^{{-q_n \gamma}}} \Psi_{\gamma} (\gamma)   d\gamma
	\end{eqnarray}
where the set $\{N, \delta, p, q_n\}$ can specify a variety of modulation schemes. 
\begin{my_lemma}[Average BER for FPT]
	\label{theorem:ber_fpt}
	If $k_i$ and $\beta_i$ are the parameters of the foggy channel,  $A_i$ and $\rho_i$ are the  parameters of pointing errors,  sets $(\alpha_i,\beta_i,\Omega_i)$, $(\phi_i,\varphi_i,\Xi_i)$ are the parameters for atmospheric turbulence, and $z_i=4.343/\beta_i^{\rm fog} d_i$, then an exact expression of  the average BER for the FPT channel is
	\begin{eqnarray}
		\bar{P}^{\rm FPT}_e=\bar{P}^{\rm FPT}_{e,1}+\bar{P}^{\rm FPT}_{e, 2}-2\bar{P}^{\rm FPT}_{e, 1}\bar{P}^{\rm FPT}_{e, 2}
	\end{eqnarray}
	where $\bar{P}^{\rm FPT}_{e,1}$ and $\bar{P}^{\rm FPT}_{e, 2}$ are given in \eqref{eq:ber_fpt} with $i=1$ and $i=2$:
			\begin{eqnarray} \label{eq:ber_fpt}
			&\bar{P}^{\rm FPT}_{e,i} = \frac{\delta z_i^{k_i}\rho_i^{2}\sigma_i^{\beta_i-\frac{1}{2}}\lambda_i^{\varphi_i-\frac{1}{2}} (2\pi)^{1-\frac{\lambda_i+\sigma_i}{2}}}{4\Gamma(p)(\phi_i\lambda_i)^{k_i}\Gamma(\beta_i)\Gamma(\varphi_i)} \sum_{n=1}^{N}\nonumber \\& H_{\lambda_i+\sigma_i+k_i+2,3+k_i}^{2,\lambda_i+\sigma_i+k_i+1} \bigg[\begin{array}{c}\frac{\lambda_i^{\lambda_i}\sigma_i^{\sigma_i}\Omega_i^{\sigma_i}\Xi_i^{\lambda_i}}{\beta_i^{\sigma_i}\varphi_i^{\lambda_i}} \big(A_{i}\sqrt{q_n\gamma_{0}}\big)^{\phi_i\lambda_i}\end{array} \bigg\vert \begin{array}{c} \tilde{U}_i\\ \tilde{V}_i \end{array}\bigg]\nonumber \\&
		\end{eqnarray}
	where $\tilde{U}_i= \{(\mu_i,1), (\{1-\frac{z_i}{\phi_i\lambda_i}\}_{1}^{k_i},1), (1,\frac{\phi_i\lambda_i}{2})\}$ and $\tilde{V}_i=(p,\frac{\phi_i\lambda_i}{2}) \{(0,\frac{\phi_i\lambda_i}{2}),(\nu_i,1), (\{-\frac{z_i}{\phi_i\lambda_i}\}_{1}^{k_i},1)\}$.
	\end{my_lemma}

\begin{IEEEproof}
 Using \eqref{eq:snr_cdf_fpt} in \eqref{eq:ber} and applying the definition of Fox H-function with inner integral $\int_{0}^{\infty}\gamma^{p-\frac{\phi_i\lambda_is}{2}-1}e^{-q_n\gamma}dt=\frac{\Gamma\left(p-\frac{\phi_i\lambda_is}{2}\right)}{q^{p-\frac{\phi_i\lambda_is}{2}}_n}$, we get 
 \begin{eqnarray} \label{ber_fpt2}
&\bar{P}^{\rm FPT}_{e,i} =\frac{\delta z_i^{k_i}\rho_i^{2}\sigma_i^{\beta_i-\frac{1}{2}}\lambda_i^{\varphi_i-\frac{1}{2}} (2\pi)^{1-\frac{\lambda_i+\sigma_i}{2}}}{4\Gamma(p)(\phi_i\lambda_i)^{k_i}\Gamma(\beta_i)\Gamma(\varphi_i)} \sum_{n=1}^{N}\nonumber\\& \frac{1}{2\pi i}\int_{L}^{}\frac{\Gamma\left(p-s\frac{\phi_i\lambda_i}{2}\right)\Gamma\left(\frac{z_i}{\phi_i\lambda_i}+s\right)\Gamma(-s\frac{\phi_i\lambda_i}{2})\prod_{j=1}^{\lambda_i+\sigma_i+1}\Gamma(1-\mu_{i,j}+s)}{\Gamma(1-\nu_i+s)\Gamma(1-s\frac{\phi_i\lambda_i}{2})\Gamma\left(1+\frac{z_i}{\phi_i\lambda_i}+s\right)}\nonumber\\&\left(\frac{\lambda_i^{\lambda_i}\sigma_i^{\sigma_i}\Omega_i^{\sigma_i}\Xi_i^{\lambda_i}}{\beta_i^{\sigma_i}\varphi_i^{\lambda_i}} \big(A_{i}\sqrt{q_n\gamma_{0}}\big)^{\phi_i\lambda_i}\right)^sds 
\end{eqnarray}   
Applying the definition of Fox-H function \cite{Kilbas}, we get the average BER in \eqref{eq:ber_fpt}. 
It is well knowm that the average BER for the DF relaying with Gray coding can also be expressed as \cite{Tsiftsis2006}:
\begin{eqnarray} 
	\label{eq:zaf5}
	\bar{P_e} = \bar{P_e}_1 + \bar{P_e}_2 - 2 \bar{P_e}_1 \bar{P_e}_2
\end{eqnarray}
where $\bar{P_e}_1$ and $\bar{P_e}_2$ denote the average BER of the first link and the second link, respectively.	
Thus,  using $i=1,2$ in \eqref{eq:ber_fpt} with \eqref{eq:zaf5}, we prove  Lemma 5.
 \end{IEEEproof}	
We  use \cite{Kilbas} to present an asymptotic expression at high SNR for the average BER
\begin{eqnarray} \label{eq:ber_asym}
	&\bar{P}^{\infty}_{e,i} =\frac{\delta z_i^{k_i}\rho_i^{2}\sigma_i^{\beta_i-\frac{1}{2}}\lambda_i^{\varphi_i-\frac{1}{2}} (2\pi)^{1-\frac{\lambda_i+\sigma_i}{2}}}{4\Gamma(p)(\phi_i\lambda_i)^{k_i}\Gamma(\beta_i)\Gamma(\varphi_i)} \sum_{n=1}^{N}\nonumber\\&\sum_{m=1}^{\lambda_i+\sigma_i+k_i+1}\Gamma\left((1-a_m)\frac{\phi_i\lambda_i}{2s_m}\right)\nonumber\\&\frac{\Gamma\left(p-(a_m-1)\frac{\phi_i\lambda_i}{2s_m}\right)\prod_{j=1,j\neq m}^{\lambda_i+\sigma_i+k_i+1}\Gamma\left(1-a_j+(a_m-1)\frac{s_j}{s_m}\right)}{s_m\Gamma\left(1-(a_m-1)\frac{\phi_i\lambda_i}{2s_m}\right)\prod_{j=2}^{k_i+2}\Gamma\left(1-b_j+(a_m-1)\frac{t_j}{s_m}\right)}\nonumber\\& \left(\frac{\lambda_i^{\lambda_i}\sigma_i^{\sigma_i}\Omega_i^{\sigma_i}\Xi_i^{\lambda_i}}{\beta_i^{\sigma_i}\varphi_i^{\lambda_i}} \big(A_{i}\sqrt{q_n\gamma_{0}}\big)^{\phi_i\lambda_i}\right)^{\frac{(a_m-1)}{s_m}}
\end{eqnarray}
where $a_m=a_j=\{\mu_i, \{1-\frac{z_i}{\phi_i\lambda_i}\}_{1}^{k_i}, 1\}$, $s_m=s_j=\{1, 1, \frac{\phi_i\lambda_i}{2}\}$, $b_m=b_j=\{0, p, \nu_i, \{-\frac{z_i}{\phi_i\lambda_i}\}_{1}^{k_i}\}$ and $t_m=t_j=\{\frac{\phi_i\lambda_i}{2}, \frac{\phi_i\lambda_i}{2}, 1, 1\}$.
Similar to the outage probability, we can use \eqref{eq:ber_asym} to obtain the diversity order of the dual-hop relay-assisted system as $M_{{\rm BER}}^{\rm FPT}= \min\{\frac{\rho_1^2}{2},\frac{\rho_2^2}{2}, \frac{\alpha_1\beta_1}{2},\frac{\phi_1\varphi_1}{2},  \frac{\alpha_2\beta_2}{2}, \frac{\phi_2\varphi_2}{2}, \frac{z_1}{2}, \frac{z_2}{2} \}$.

Next, we analyze the average BER performance for the FP channel. However, general solution to the integration in \eqref{eq:ber} with \eqref{eq:snr_cdf_3}  is intractable due to the presence of exponential function and incomplete Gamma function with logarithmic argument raised to the power $k$ in the CDF function for the FP channel. Thus, we derive a closed-form expression for the average BER for a particular integer value of $k$ (i.e., $k=2$, $k=3$, and so on). Specifically, we consider light foggy condition (with a shape parameter $k = 2$) and  general pointing errors  with asymptotic atmospheric turbulence  to provide exact closed-form expression on the average BER. 
\begin{my_lemma}[Average BER for FP]
\label{lemma:ber_fp}
	If $k=2$ and $\beta^{\rm{fog}}$ are  the parameters of the foggy channel,  $A_1$, $A_2$ and $\rho_1$, $\rho_2$ are the parameters of pointing errors and atmospheric turbulence, and $z_1=4.343/\beta^{\rm{fog}} d_1$, $z_2=4.343/\beta^{\rm{fog}} d_2$ with $d_2\geq d_1$, then a closed-form expression of average BER for the FP channel is:
	\begin{eqnarray}
		\bar{P}^{\rm FP}_e=\bar{P}^{\rm FP}_{e,1}+\bar{P}^{\rm FP}_{e,2}-2\bar{P}^{\rm FP}_{e, 1}\bar{P}^{\rm FP}_{e,2}
	\label{eq:asymt_ber}
	\end{eqnarray}
where $\bar{P}^{\rm FP}_{e,1}$ and $\bar{P}^{\rm FP}_{e, 2}$ are given in \eqref{eq:avg_ber_fp} with $i=1$ and $i=2$.
\end{my_lemma}
	\begin{flalign}
	&\bar{P}^{\rm FP}_{e,i} =\frac{\delta}{2\Gamma(p)}\sum_{j=1}^{N}q^{p}_j\Bigg[\frac{ D^{(1)}(A_i\sqrt{\gamma_0})^{-\rho_i^2}}{q^{p+\frac{\rho^2_{i}}{2}}_j}\big(\Gamma\big(p+\frac{\rho_i^2}{2}\big)\nonumber\\&-\Gamma\big(p+\frac{\rho_i^2}{2}, q_jA_i\gamma_0\big)\big)-D^{(2)}\Big[\frac{(A_i\sqrt{\gamma_0})^{-z_i}}{z_iq^{p+\frac{z_{i}}{2}}_j}\big(\Gamma\big(p+\frac{z_i^2}{2}\big)\nonumber\\&-\Gamma\big(p+\frac{z_i^2}{2}, q_jA^2 _i\gamma_0\big)\big) \big(1+\frac{m_i}{z_i}\big)+\frac{m_i(A_i\sqrt{\gamma_0})^{-(z_i+1)}}{2z_iq^{p+\frac{z_{i}}{2}+1}_j}\nonumber\\&\Big(G_{2,3}^{3,0} \bigg[\begin{array}{c}q_jA^2_i\gamma_0\end{array} \bigg\vert \begin{array}{c} 1,1 \\ 0,0,\frac{2p+z_i+1}{2} \end{array}\bigg]\nonumber\\&-\Gamma\big(\frac{2p+z_i+1}{2}\big)\psi^{(0)}\big(\frac{2p+z_i+1}{2}\big)\Big)\Big]\Bigg].
		\label{eq:avg_ber_fp}
	\end{flalign}

\begin{IEEEproof} 
First, we substitute \eqref{eq:aprox_snr_cdf} and  \eqref{eq:snr_cdf_3} in \eqref{eq:ber}.  We use $m_1=(z_1-\rho_1^2)$, $m_2=(z_2-\rho_2^2)$ and  the series expansion $\Gamma(a,t) \triangleq (a-1)! e^{-t} \sum_{m=0}^{a-1} \frac{t^m}{m!}$, and substitute $u=\frac{A_1}{\sqrt{\gamma/\gamma_0}}$, and $v=\frac{A_2}{\sqrt{\gamma/\gamma_0}}$ to simplify the integrations into  algebraic functions. Apart from simple integration terms, we also encounter the following  terms:
\begin{eqnarray}
&\int_1^{\infty } u^{-n-2} e^{-\frac{p}{u^2}} du, \int_{a}^{\infty }u^{-n-2} e^{-\frac{p}{u^2}} \ln(u)  du
\label{eq:identity_1_asmyt_ber1}
\end{eqnarray}
 We provide  closed-form expressions of \eqref{eq:identity_1_asmyt_ber1} in Appendix B.  Using \eqref{eq:identity_1_asmyt_ber2}, \eqref{eq:identity_3_asmyt_ber2} of Appendix B with some algebraic simplifications, we get \eqref{eq:avg_ber_fp}. Finally, using $i=1, 2$ in \eqref{eq:avg_ber_fp} with \eqref{eq:zaf5}, we prove  Lemma 6.
\end{IEEEproof} 

Similarly, the  average BER expression for other values of $k$ can be obtained. It should be noted that the use of PDF to derive the average BER results into more terms compared to the CDF based approach of \eqref{eq:ber}. The average BER in 	\eqref{eq:avg_ber_fp} is useful in analyzing the system performance using known mathematical functions. To simplify the underlying expressions further, we consider  deterministic path gain in the following Corollary.

\begin{my_corollary}[Average BER for FP with deterministic path gain]
	\label{proposition_ber_asymptotic}
	If the fog  causes a deterministic path gain  $L_r= e^{-\tau d/2} $ with relay-assisted transmission  $d_1=d_2= d/2$ for an i.i.d fading model  	under random pointing errors and then expression of average BER is given as
		\begin{flalign}
	&\bar{P}^{\rm FP}_{e,i}= \sum_{n=1}^{N}\frac{\delta}{\Gamma(p)(AL_r\sqrt{q_n\gamma_{0}})^{\rho^2}}\nonumber\\&\Big[\Gamma\big(\frac{\rho^2}{2}+p\big)-\Gamma\big(\frac{\rho^2}{2}+p,q_n(AL_r)^2\gamma_{0}\big)\nonumber\\&-\frac{\Gamma\big(\rho^2+p\big)+\Gamma\big(\rho^2+p,q_n(AL_r)^2\gamma_{0}\big)}{2(AL_r\sqrt{q_n\gamma_{0}})^{\rho^2}}\Big]
	\label{eq:ber_asymp}
	\end{flalign}
\end{my_corollary}

\begin{IEEEproof} 
	Using \eqref{eq:aprox_snr_cdf} in  \eqref{eq:ber}, and noting that $A_1=A_2=A$, and  $\rho_1=\rho_2=\rho$ under the symmetric i.i.d fading model, we get:
	\small
	\begin{align}
	\begin{split}
	\bar{P}^{\rm FP}_{e,i}=\frac{\delta}{2\Gamma(p)}\sum_{n=1}^{N}q^p_n\int_{0}^{ (AL_r)^2\gamma_0}\gamma^{p-1}e^{-q_n\gamma}[2F_{\gamma}(\gamma)-(F_{\gamma}(\gamma))^2]d\gamma
	\label{func:symt_ber_tp}
	\end{split}
	\end{align}
	Using \eqref{eq:pdf_SNR_atm_point_asymp} and substituting $F_\gamma(\gamma)=\int_{0}^{\gamma}f(\gamma)d\gamma$  in \eqref{func:symt_ber_tp} and applying the definition of Gamma function and incomplete Gamma function, we get \eqref{eq:ber_asymp}.
\end{IEEEproof}  
Similar to the average SNR and ergodic capacity, it can be seen  from	\eqref{eq:ber_asymp} that the consideration of  deterministic path gain simplifies the  BER expression compared with the random path gain due  to the fog.

\begin{table}[tp]	
	\renewcommand{\arraystretch}{01}
	\caption{Simulation Parameters}
	\label{table:simulation_parameters}
	\centering
	\begin{tabular}{|c|p{1cm}|p{2.5cm}|}
		\hline 	
		Transmitted power &$P_t$ & $0$ to $40$ \mbox{dBm} \\ \hline
		Responsitivity &	$R$ & $0.5$ \mbox{A/W}\\ \hline
		
		AWGN variance &$\sigma_w^2$ & $10^{-14}~\rm {A^2/GHz}$\footnotemark{} \\ \hline	
		Link distance &$d$ & $400$ \mbox{m} and $1200$ \mbox{m}\\ \hline	
		Shape parameter of fog &$k$ & \{2.32, 5.49, 6.00\} \\ \hline	
		Scale parameter of fog &$\beta$ & \{13.12, 12.06, 23.00\}\\ \hline Aperture diameter & $D=2a_r$ & $10$ \mbox{cm} \\ \hline Normalized beam-width &$w_z/a_r$ & \{15, 20, 25\}\\ \hline Normalized jitter &$\sigma_{s}/a_r$ & \{3, 5\} \\ \hline Refractive index&$C_n^2$ & $8 \times10^{-14}$\\ \hline Wavelength &$\lambda$ & $1550$ \mbox{nm} \\ \hline  Turbulence parameters &$\{\alpha, \beta, \omega\}$ $\{\phi, \varphi, \Xi\}$  & \cite{Kashani2015}\\ \hline
	\end{tabular}	 
\end{table} 
\footnotetext[1]{In many papers, it is wrongly written as $\rm A^2/Hz$.}

\begin{figure*}[tp]
	\begin{center}
		\subfigure[Symmetrical: $d_1=d_2=400$ with $w_z/a_r=25$, $\sigma_{s}/a_r=3$ \mbox{m}.]{\includegraphics[width=\columnwidth]{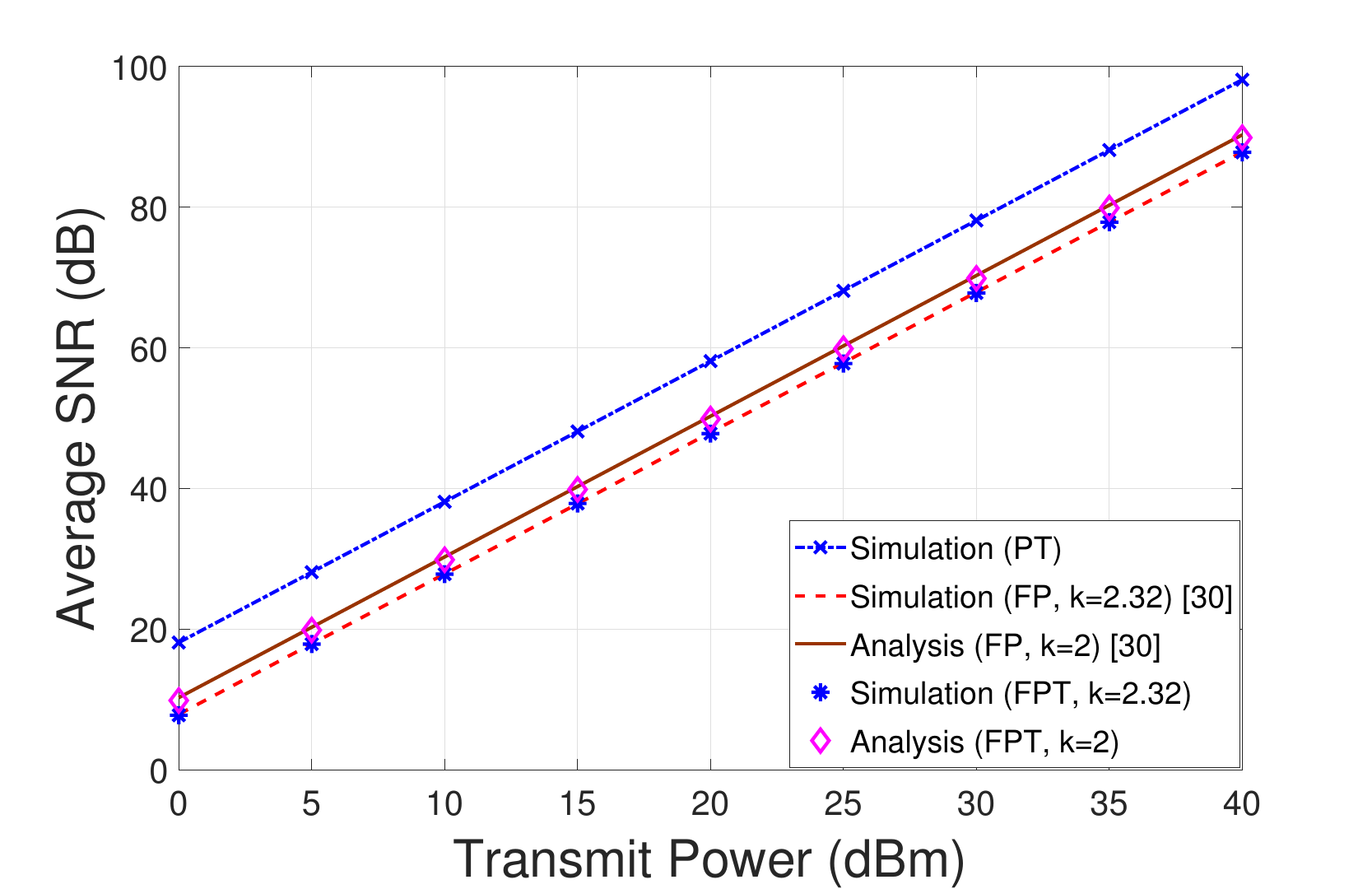}}\quad
		\subfigure[Asymmetrical: $d_1=300$ \mbox{m}, $d_2=500$ \mbox{m}]{\includegraphics[width=\columnwidth]{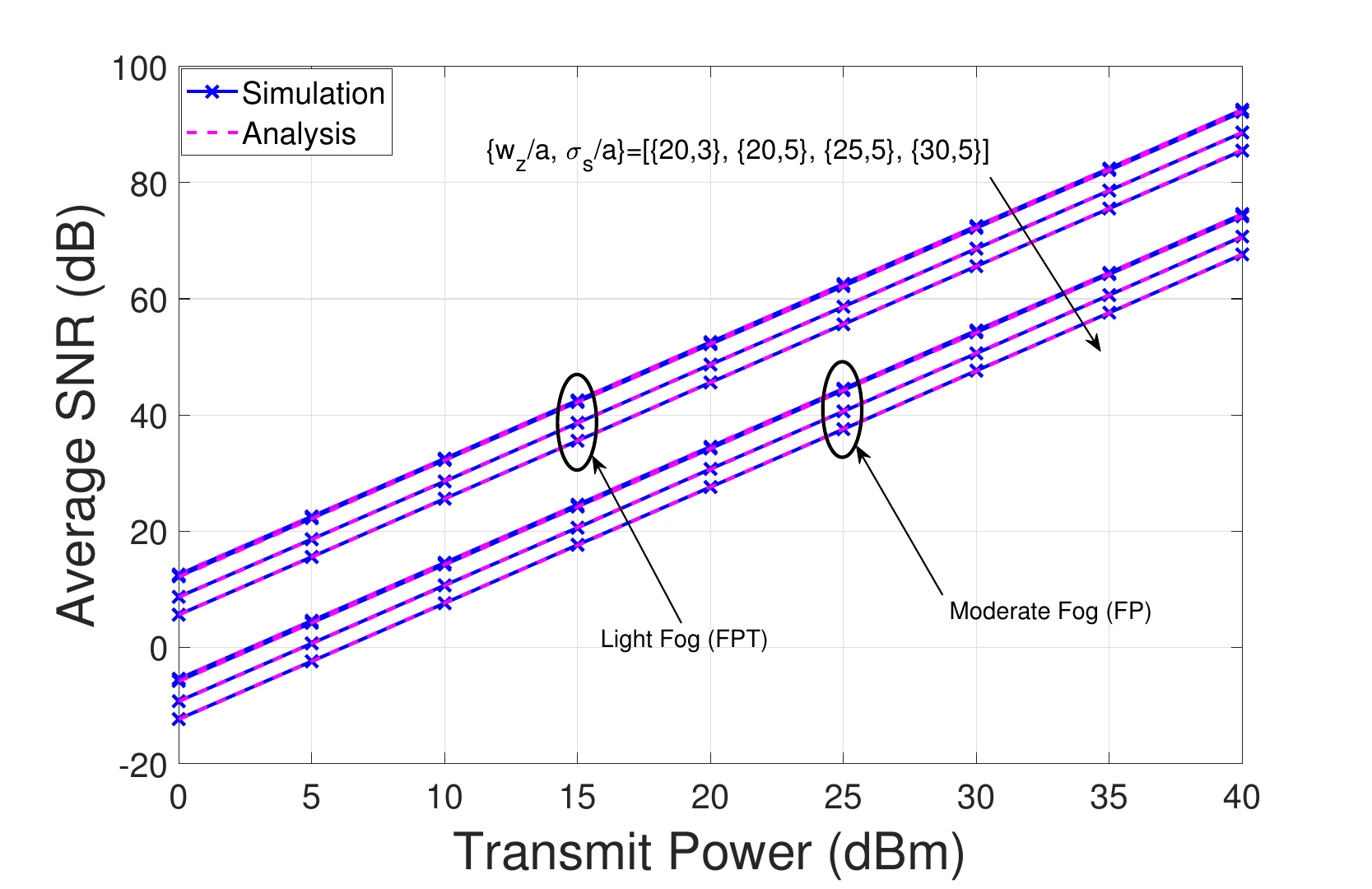}}
		\caption{Average SNR performance of relay-assisted OWC system for a transmission link $d=800$ \mbox{m}.}
		\label{snr_d_600m_snr_d_1}
	\end{center}
\end{figure*}

\begin{figure*}[tp]
	\begin{center}
		\subfigure[FP channel for different fog conditions at $\sigma_{s}/a_r=3$ and $w_z/a_r=25$, asymmetrical: $d_1=200$ \mbox{m}, $d_2=300$ \mbox{m}, and symmetrical: $d_1=d_2=250$ \mbox{m}.]{\includegraphics[width=\columnwidth]{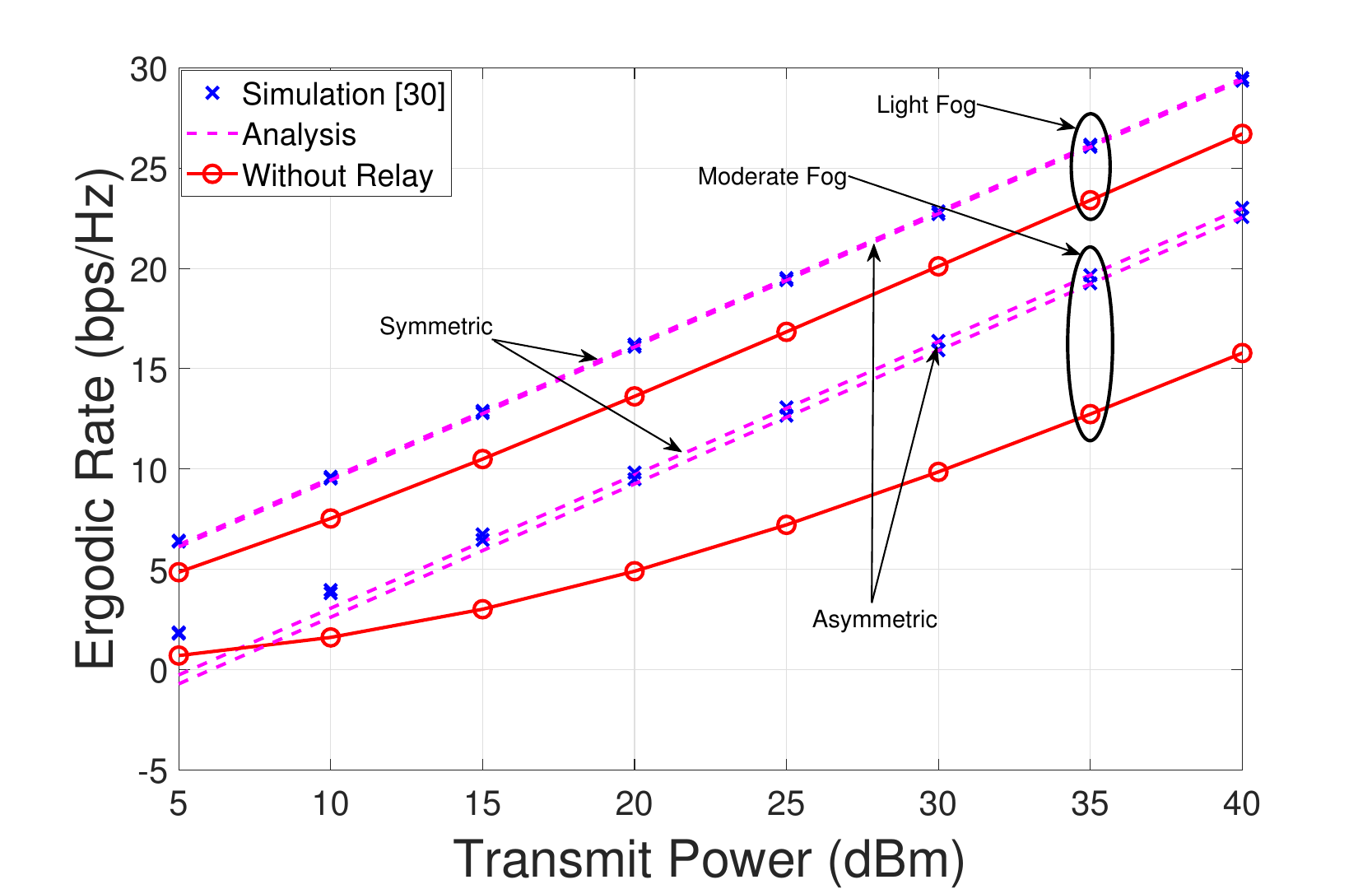}}\quad
		\subfigure[FPT channel with light fog and strong turbulence for  different $w_z/a_r$ at  $\sigma_{s}/a_r=3$  and  asymmetrical  relay position: $d_1=300$ \mbox{m}, $d_2=500$ \mbox{m}.]{\includegraphics[width=\columnwidth]{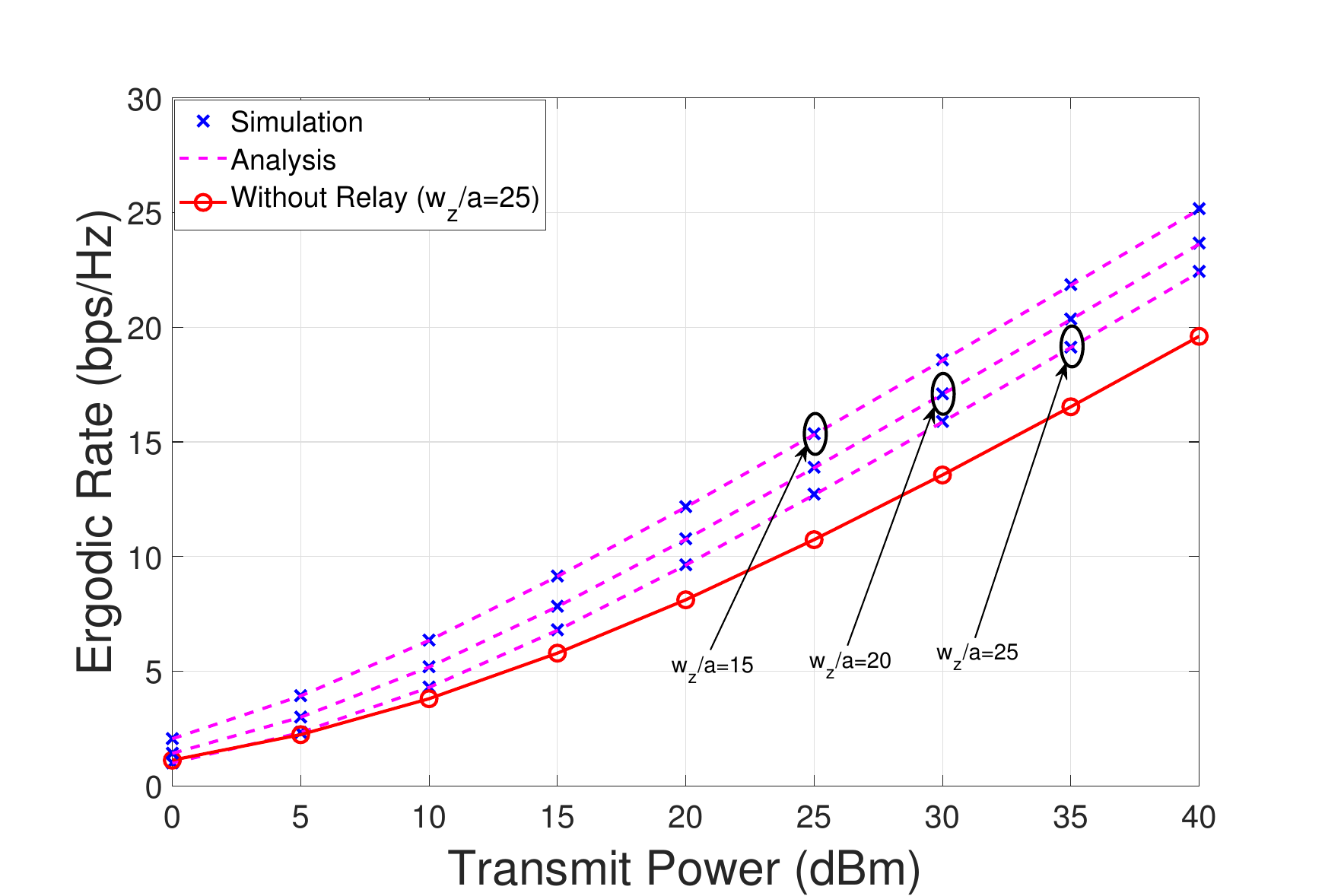}}
		\caption{Ergodic rate performance of relay-assisted OWC system.}
		\label{rate_perf}
	\end{center}
\end{figure*}
\section{Simulation and Numerical Analysis}
In this section, we use numerical analysis and Monte Carlo (MC) simulation (averaged over $10^7$ channel realizations) to demonstrate the performance of the relay-assisted OWC system  under the combined effect of fog with pointing errors.  We consider three simulation scenarios: fog, pointing errors, and atmospheric turbulence (FPT), fog and pointing errors (FP), deterministic path gain with pointing errors, and atmospheric turbulence (PT). We also compare the performance of the FP channel with the state-of-the-art paper \cite{Esmail2017_Photonics}.   We  validate our derived analytical expressions with numerical and simulation results.   Although a direct link between the source and destination may not exist, we also compare the performance of direct and relay-assisted transmissions at various link distances  and pointing errors parameters.  We  consider  strong ($\alpha=1.8621$, $\beta=0.5$, $\Omega=1.5074$, $\phi=1$, $\varphi=1.8$, and $\Xi=0.928$), moderate ($\alpha=2.169$, $\beta=0.55$, $\Omega=1.5793$, $\phi=1$, $\varphi=2.35$, and $\Xi=0.9671$), and weak ($\alpha=2.1$, $\beta=4$, $\Omega=1.0676$, $\phi=2.1$, $\varphi=4.5$, and $\Xi=1.06$) turbulence conditions to model the DGG atmospheric turbulence \cite{Kashani2015}. We use $\lambda=28, \sigma=13$ for strong turbulence and $\lambda=17, \sigma=9$ \cite{AlQuwaiee2015} for moderate turbulence to compute analytical expressions.  We use other standard simulation parameters of the OWC system  as given in Table \ref{table:simulation_parameters}.

First, we demonstrate the mutual effects of different channel impairments on the OWC system by plotting the average SNR versus transmit power for a link distance of $800$ \mbox{m}, as shown in     Fig.~\ref{snr_d_600m_snr_d_1}. Comparing the PT (where deterministic path gain is considered) and FPT plots in Fig.~\ref{snr_d_600m_snr_d_1}(a), we can find that the fixed visibility range based path-gain computation overestimate the average SNR by $10$ \mbox{dB} with respect to the random path gain consideration. Further, the OWC system performs similarly for FP and FPT channels, as shown in  Fig.~\ref{snr_d_600m_snr_d_1}(a), since the effect of fading due to the atmospheric turbulence is negligible in the presence  of fog for shorter links.  Moreover, it can also be seen that there is a small difference of around $2$ \mbox{dB} in the average SNR  between the proposed analysis for FP/FPT channels (which is based on integer-valued $k=2$) with simulation results on real-valued $k=2.32$ for light fog.  In Fig.~\ref{snr_d_600m_snr_d_1}(b), we analyze the effect of pointing errors on the average SNR performance over  light fog (with strong turbulence) under the FPT channel and moderate fog under the FP channel considering the asymmetric placement of the relay ($d_1=300$ \mbox{m} and $d_2=500$ \mbox{m}).  It can be seen from the figure that the average SNR performance decreases with an increase in normalized beam width and jitter.  It should be noted that the effect of normalized beam width $(w_z/a_r)$ on the average SNR performance is more as compared to the  standard deviation of the normalized jitter  $(\sigma_{s}/a_r)$.

The ergodic rate performance in Fig.~\ref{rate_perf} shows  a significant benefit of the relay-assisted system under  FP and FPT channels. However, similar to the average SNR, the impact of atmospheric turbulence is found to be negligible on the ergodic rate in the presence of  fog. The relay-assisted system gives more increment  over moderate fog than the light fog as compared to the no-relay system. Moreover, the slopes  at high transmit power  show  greater improvement with the relay-assisted transmission than with direct transmissions.  Comparing Fig.~\ref{rate_perf}(a) and  Fig.~\ref{rate_perf}(b), it can be seen a greater improvement in the ergodic rate performance using the asymmetric placement of the relay with a longer link length (i.e., $800$ \mbox{m}) than a shorter one (i.e., $500$ \mbox{m}). It can also be seen from  Fig.~\ref{rate_perf}(b) that the normalized beam width has a significant impact on the ergodic rate performance. 
 
\begin{figure*}[tp]
  		\begin{center}
  			\subfigure[FP channel with light fog (top) and moderate fog (bottom).]{\includegraphics[width=\columnwidth]{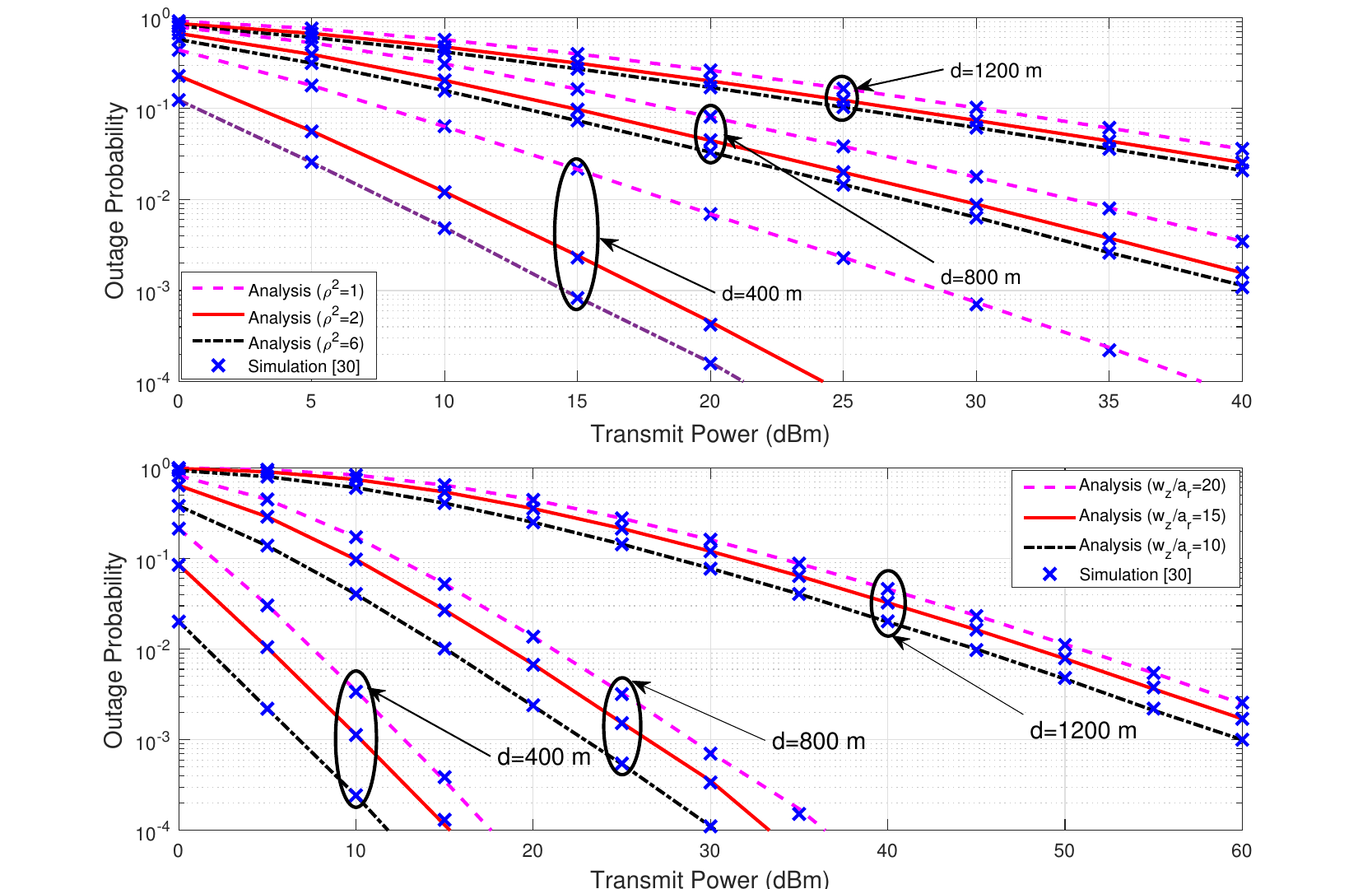}}\quad
  			\subfigure[FPT channel with light fog and pointing error parameters for the first link with  $\sigma_{s}/a_r=3$ and $w_z/a_r=8$, and the second link with  $\sigma_{s}/a_r=6$ and $w_z/a_r=25$.]{\includegraphics[width=\columnwidth]{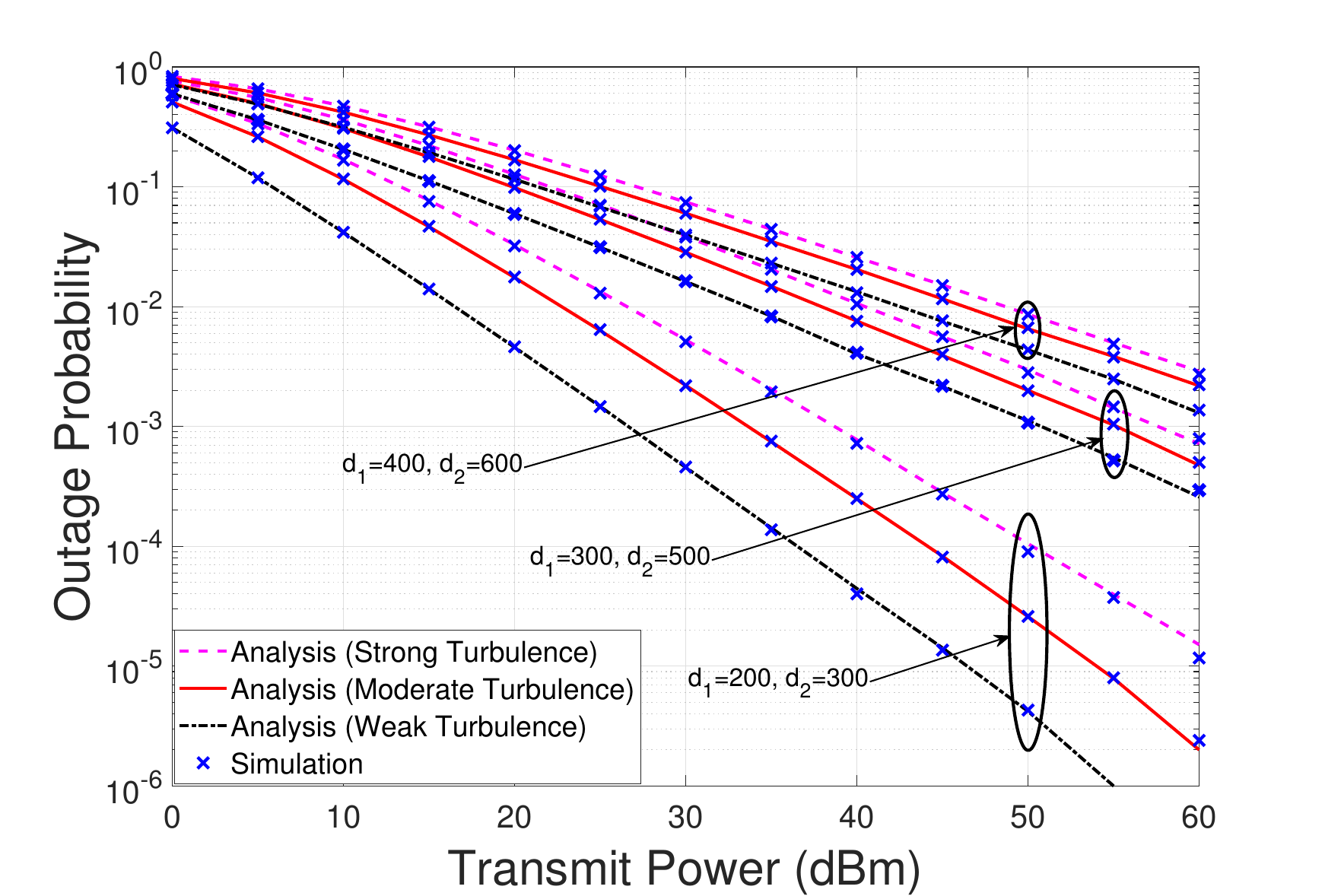}}
  			\caption{Outage performance of relay-assisted OWC system at a threshold SNR $\gamma_{\rm th} = 6$ \mbox{dB}.}
  			\label{out_perf}
  		\end{center}
  	\end{figure*}

 In Fig.~\ref{out_perf}(a), we demonstrate the outage probability performance of the OWC system for the FP channel with two  foggy conditions (i.e., light and moderate) with different pointing errors conditions and symmetric link distances.  For the moderate fog, the normalized jitter of the  pointing errors is   $\sigma_s/r=3$. We  compare  the two subplots of the figure to show that an increase in the fog density deteriorates the outage probability performance of the OWC system: transmission in moderate fog requires almost $16$\mbox{dBm} more transmit power to achieve the same outage probability $10^{-4}$  for a link distance of $400$\mbox{m}  in the light fog condition.  Further, Fig.~\ref{out_perf}(a) shows that communication range with the light fog is limited to $800$\mbox{m} at a transmit power of $40$\mbox{dBm} to achieve an acceptable outage probability of $10^{-3}$.
 
  We consider the channel and system parameters judicially to demonstrate the diversity order of the system. For the light foggy condition having parameter $\beta^{\rm fog}=13.12$, the diversity order for the $1200$\mbox{m} link is $M_{\rm out}^{\rm FP}=\min\{ \frac{z_i}{2}, \frac{\rho_i^2}{2}\}= 0.27$ for each pointing error parameter $\rho^2=1,2, 6$. Similarly, the diversity order for the $800$\mbox{m} link is $M_{\rm out}^{\rm FP}= 0.42$ for each $\rho^2=1,2, 6$.  Thus, the diversity order becomes independent of the pointing error parameter $\rho^2$, which can be confirmed through  the slope of plots for varying pointing error parameters for both $1200$\mbox{m} and $800$\mbox{m} link distances. However, the diversity order for the $400$\mbox{m} link is  $M_{\rm out}^{\rm FP}= 0.5$ with $\rho^2=1$ (limited by pointing errors)   and $M_{\rm out}^{\rm FP}= 0.83$ with $\rho^2=2, 6$ (limited by fog). The slope of plots for the $400$\mbox{m} link confirms the diversity order behavior.    Comparing the plots for $1200$\mbox{m},  $800$\mbox{m}, and $400$\mbox{m},  it can be seen from Fig.~\ref{out_perf}(a) that there is a change in the slope since the diversity order is different for the considered  link distances. Similar observations can be inferred  for the moderate fog with $\beta^{\rm fog}=12.06$ (see the second subplot of Fig.~\ref{out_perf}(a)). However, for the moderate fog even for the $400$\mbox{m} link, the diversity order is dependent on the foggy condition since the minimum value of pointing error parameter $\rho^2=2.84$ (computed from $\sigma_{s}/a_r=3$ and $w_z/a_r=10$) is greater than $z=1.6551$.
  		
In Fig.~\ref{out_perf}(b), we demonstrate the performance of outage probability  for the FPT channel for light  foggy conditions with different pointing errors and turbulence conditions for asymmetric link distances.   It can be seen that the intensity of fluctuations has an impact on the outage probability. For a link distance of $1000$\mbox{m}, almost $8$\mbox{dBm} higher transmit power is required to achieve the same probability for the strong turbulence when compared to weak turbulence at a transmit power of $60$\mbox{dBm}.     Similar to the FP channel,  slope of  plots demonstrate the diversity order of the system.  Note that  strong,  moderate, and weak turbulence can introduce a diversity order of $0.47$, $0.60$, $4.2$, respectively. Considering the link distances, the diversity order (dominated by the fog parameter) can be $0.27$ (for $d_2=600$\mbox{m}), $0.33$ (for $d_2=500$\mbox{m}), and  $0.55$ (for $d_2=300$\mbox{m}). Further, the diversity order from  pointing errors can be $0.5878$ using the parameters $\sigma_{s}/a_r=3$ and $w_z/a_r=8$. Thus, using  $ M_{\rm out}^{\rm FPT}=\min\{\frac{z_i}{2}, \frac{\rho_i^2}{2}, \frac{\alpha_i\beta_i}{2}, \frac{\phi_i\varphi_i}{2}\}$, the diversity order of the FPT channel is $0.27$ (for the link $1000$\mbox{m}) and $0.33$ (for the link $800$\mbox{m}). However, for the $500$\mbox{m}, the diversity order is $0.47$ as determined from the strong turbulence. It can be seen that the slope of plots in Fig.~\ref{out_perf}(b) demonstrates the diversity order behavior. When comparing the plots for shorter and longer links, the diversity order depends on  the fog parameters for longer links and may depend on the pointing errors and turbulence for shorter links.

 \begin{figure*}[tp]
 	\begin{center}
 		\subfigure[FP channel with light fog (top) and moderate fog (bottom with $\sigma_{s}/a_r=3$).]{\includegraphics[width=\columnwidth]{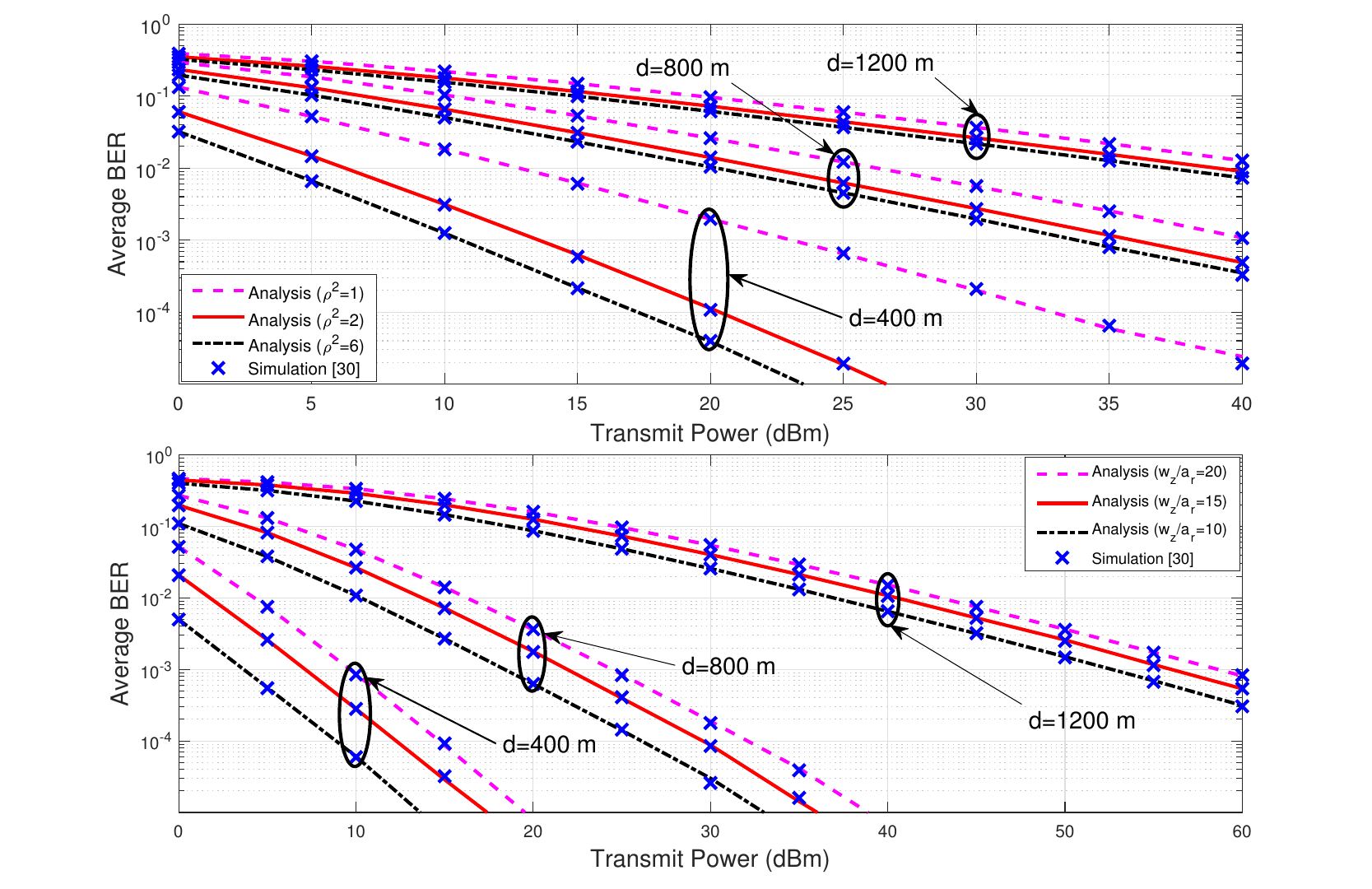}}\quad
 		\subfigure[FPT channel with pointing error parameters for the first link  $\sigma_{s}/a_r=3$ and $w_z/a_r=6$ and the second link with  $\sigma_{s}/a_r=6$ and $w_z/a_r=25$. ]{\includegraphics[width=\columnwidth]{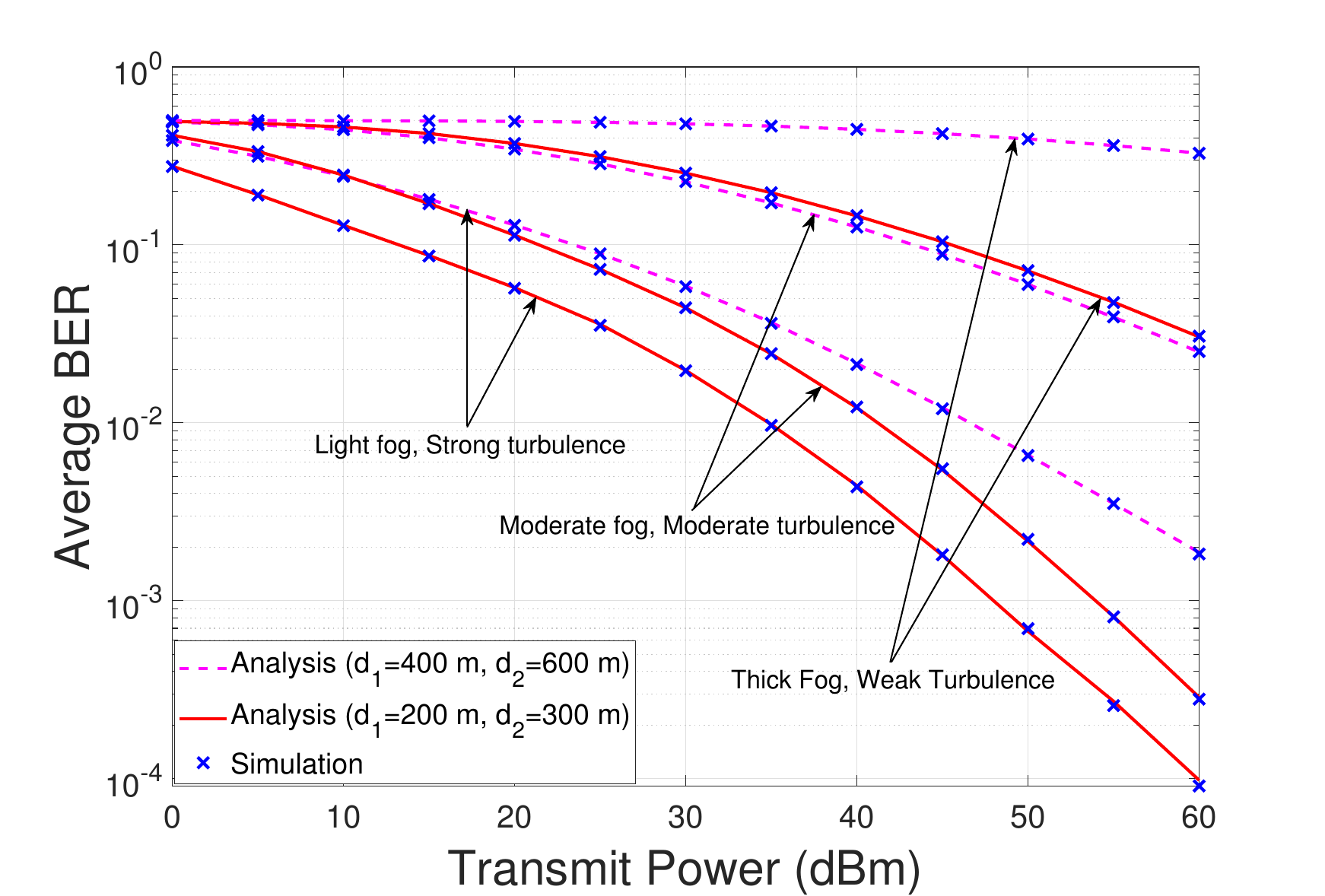}}
 		\caption{Average BER performance of relay-assisted OWC system.}
 		\label{ber_perf}
 	\end{center}
 \end{figure*}

In Fig.~\ref{ber_perf}, we demonstrate the average BER performance of the OWC system over the FP channel with the popular on-off keying  (OOK) modulation using $\delta=1$, $N=1$, $p=1/2$, and  $q=1/2$ as a function of transmit power. In Fig.~\ref{ber_perf}(a), we consider  two  foggy conditions (i.e., light and moderate) with different pointing errors conditions and symmetric link distances.   Similar to the outage probability, an increase in  fog density deteriorates the average BER performance. 	The diversity order for the BER can be illustrated for the  moderate  fog condition (see the second subplot  Fig.~\ref{ber_perf}(a)). Using  $\beta^{\rm fog}=12.06$, we get $z_1=z_2= 0.60$ for the link distance $d=1200$\mbox{m}, $z_1=z_2= 0.90$ for the link distance $d=800$\mbox{m},   and $z_1=z_2= 1.80$ for the link distance $d=400$\mbox{m}. Using $\min\{z_1/2, \rho_1^2/2\}$,   the diversity order $0.30$ depends on  fog parameters for link distances $1200$\mbox{m} and $800$\mbox{m} for each pointing error parameter $\rho^2=1, 2, 6$. However, for the link distance $400$\mbox{m} and $\rho^2=1$, the diversity order $0.5$ depends on the pointing error parameter $\rho$. We can observe the slope of plots (see the second subplot of Fig.~\ref{out_perf}(a))  to confirm the behavior of diversity with different system and channel parameters. Similar observations can be inferred   from the average BER for the light fog with $\beta^{\rm fog}=13.12$ (see the first subplot of Fig.~\ref{out_perf}(a)).

 Finally, we consider a more practical situation for performance evaluation in Fig.~\ref{ber_perf}(b). Specifically, we consider three foggy conditions  light, moderate, and thick, which are each associated with corresponding atmospheric turbulence of  strong, moderate, and weak, respectively. This is a typical scenario since the atmospheric turbulence and foggy conditions are inversely correlated \cite{Farid2007}. We demonstrate the performance of average BER for the considered scenario with different foggy, pointing errors and turbulence conditions for asymmetric link distances.   It can be seen that the intensity of fluctuations has an impact on the average BER. The figure shows that the impact of random fog is more pronounced on the performance than the intensity of turbulence. It can be seen that the range for OWC link is limited  to $500$\mbox{m} for thick fog conditions. The performance under light fog with strong turbulence  is acceptable even for longer links. For a link distance of $500$\mbox{m}, almost $5$ \mbox{dBm} higher transmit power is required to achieve the same BER when the moderate fog  (with moderate turbulence) is compared with the light fog  (with strong turbulence) at a transmit power of $60$\mbox{dBm}.  As expected, the moderate fog performs in between the light fog and thick fog. Further, the change in the slope  	of plots for various channel parameters demonstrates the system's diversity order.

In all the above plots (Fig.~\ref{snr_d_600m_snr_d_1} to  Fig.~\ref{ber_perf}), we have also verified our derived expressions with the simulation and numerical results. We use MATLAB functions  to compute the Meijer-G and Fox-H functions involved in analytical expressions. It can be seen that the derived  analytical expressions of the outage probability, average SNR, ergodic rate, and average BER have an  excellent match with the simulation results. However, there is a gap between simulation and analytical results for the ergodic rate at a lower transmit power due to the use of inequality $\log_2(\gamma)\leq \log_2(1+\gamma)$. 
 
\section{Conclusions and Future Work}

In this paper, we investigated the performance of a DF  relaying based OWC system under the combined effect of  random fog,  pointing errors, and the DGG atmospheric turbulence. We have shown that the dual-hop relaying is capable of mitigating fog,  pointing errors,  and turbulence-induced fading for high-speed OWC links. We provided design criteria   of mitigating the impact of pointing errors and random fog by adjusting the beam-width and limiting the communication range.
  We analyzed the distribution functions of SNR and provided a detailed analysis for the outage probability, average SNR, ergodic rate, and average BER  considering a generalized i.ni.d fading model  with asymmetrical distance for relaying between the source and destination. Using the asymptotic analysis, we have shown that the diversity order depends on  the fog parameters for longer links and may depend on the pointing errors and turbulence for shorter links. We have shown that there is a significant gap in the performance using the existing visibility range based path-gain computation as compared to the  probabilistic modeling of fog. Further,  the effect of normalized beam width  on the OWC performance is more than the standard deviation of  normalized jitter.    It was also demonstrated that the relay-assisted system shows better performance than the direct-link transmissions as a benchmark,  providing a more significant benefit in the denser fog, for longer link lengths, and when the relay is located symmetrically between the  source and destination.  Numerical analysis  shows that the derived closed-form expressions excellently match with simulation results, and thus can be implemented for real-time tuning of the system parameters to  optimize the OWC performance.  We envision that the consideration of the general fading model would be helpful to assess the deployment of OWC system for terrestrial wireless communications under various channel impairments.
  
 The proposed work can be augmented with several research directions.  Performance analysis for different system configurations such as amplify-and-forward in  dual-hop and multi-hop frameworks would be interesting. Further, the hybrid OWC-RF and multi-aperture systems can be investigated with the generalized fading considered in this paper.      It would also be interesting to investigate low complexity all-optical relaying schemes for OWC systems under the combined effect of fog, pointing errors, and atmospheric turbulence. 
   
\begin{appendices}
\section{(Integrals of Lemma \ref{theorem:snr_fp})}
To solve the first and second integrals of \eqref{eq:identity_1_asmyt2}, we substitute $\ln u=t$  and apply the definition of Gamma function and incomplete gamma function to get the following closed-form expressions:
\begin{eqnarray}\label{eq:identity_1_asmyt}
&	\int\limits_{1}^{\infty} u^{-n}\left(\ln(u)\right)^{p} du=\frac{\Gamma[p+1]}{(n-1)^{p+1}}
\end{eqnarray}
\begin{eqnarray}
\int\limits_{a}^{\infty} u^{-n-3}\left(\ln(u)\right)^{p} du=\frac{\Gamma[p+1,(2+n)\ln(a)]}{(2+n)^{p+1}}
\label{eq:identity_2_asmyt}
\end{eqnarray}

Further, to solve the third integral of \eqref{eq:identity_1_asmyt2}, we substitute $n\ln u=t$ and use the well-known identity $\int_{0}^{\infty}e^{-at}\Gamma(b,t)dt=a^{-1}\Gamma(b)(1-(a+1)^{-b})$ [\cite{Zwillinger2014}--pp.657, eq. 6.451.2] to get the following:
\begin{eqnarray}
 \int\limits_{1}^{\infty} u^{-n} \Gamma[k,n\ln u] du=\frac{\left(1-n^k(2n-1)^{-k}\right)\Gamma[k]}{n-1}
\label{eq:identity_3_asmyt}
\end{eqnarray}
Finally, we solve \eqref{eq:identity_4_asmyt2} by substituting $\ln u =t$ and $\ln a+t=v$ and apply the identity $\int_{u}^{\infty} x^{v-1} (x-u)^{\mu-1} e^{-bx} dx=b^{-\frac{\mu+v}{2}}u^{\frac{\mu+v-2}{2}}\Gamma(\mu)e^{-\frac{bu}{2}}W_{\frac{v-\mu}{2},\frac{1-\mu-v}{2}}(bu)$ [\cite{Zwillinger2014}--pp.348, eq. 3.383.4] to get a closed-form expression in terms of Whittaker function, $W_{\lambda,\mu}(z)$ defined as:
\begin{align}
\begin{split}
W_{\lambda,\mu}(z)=\frac{\Gamma(-2\mu)}{\Gamma(\frac{1}{2}-\mu-\lambda)}M_{\lambda,\mu}(z)+\frac{\Gamma(2\mu)}{\Gamma(\frac{1}{2}+\mu-\lambda)}M_{\lambda,-\mu}(z)
\label{eq:whittaker_function}
\end{split}
\end{align}
where $M_{\lambda,\mu}(z)$ denotes the Whittaker $M$-function. Now,  we express the Whittaker  $M$-function in terms of confluent hypergeometric function [\cite{Zwillinger2014}--pp.1024, eq. 9.220.4]
 $M_{\lambda,\mu}=z^{\mu+\frac{1}{2}}e^{-z/2}\Phi(\mu-\lambda+\frac{1}{2},2\mu+1:z)$, $M_{\lambda,-\mu}=z^{-\mu+\frac{1}{2}}e^{-z/2}\Phi(-\mu-\lambda+\frac{1}{2},-2\mu+1:z)$, where $\Phi(\alpha,\lambda;z)={}_1F_1(\alpha;\lambda;z)$, we get the following:
\small
\begin{eqnarray}
&\int\limits_{1}^{\infty} u^{-n}\left(\ln(au)\right)^{p} \left(\ln(u)\right)^{t} du=\frac{\Gamma[1+p+t]{}_1F_1(-p;-p-t;(n-1)\ln(a))}{(n-1)^{1+p+t}}\nonumber\\&+\frac{\Gamma[-1-p-t]\Gamma[1+t]\left(\ln(a)\right)^{1+p+t}{}_1F_1(1+t;2+p+t;(n-1)\ln(a))}{\Gamma[-p]}
\label{eq:identity_4_asmyt}
\end{eqnarray}
\section{(Integrals of Lemma \ref{lemma:ber_fp})}
To solve the first integral of \eqref{eq:identity_1_asmyt_ber1}, we use $\int_1^{\infty}f(t)dt=\int_0^{\infty}f(t)dt-\int_0^{1}f(t)dt$,   and apply the definition of Gamma function and  incomplete Gamma function. After a few simplifications,  we get the following closed-form expression:  
\small
\begin{eqnarray}
&\int_1^{\infty } u^{-n-2} e^{-\frac{p}{u^2}} du=\frac{1}{2} \left(\frac{1}{p}\right)^{\frac{n+1}{2}} \left(\Gamma \left(\frac{n+1}{2}\right)-\Gamma \left(\frac{n+1}{2},p\right)\right)\nonumber\\&
\label{eq:identity_1_asmyt_ber2}
\end{eqnarray}	

For the second integral of \eqref{eq:identity_1_asmyt_ber1}, we again use $\int_1^{\infty}f(t)dt=\int_0^{\infty}f(t)dt-\int_0^{a}f(t)dt$, and  apply the integration by parts on both the terms. Using the  identity $\int_0^{\infty } t^{v-1} e^{-ut} \ln (t)  dt=u^{-v} \Gamma (v) (\psi ^{(0)}(v)-\ln (u))$ [\cite{Zwillinger2014}--pp.573, eq. 4.352.1], and the  identity of Meijer's G function [\cite{Wolfram}, eq. (07.34.21.0084.01)],  we get the following:
\begin{eqnarray}
&\int_{a}^{\infty }u^{-n-2} e^{-\frac{p}{u^2}} \ln(u)  du=\frac{1}{4} p^{\frac{1}{2} (-n-1)} \nonumber\\&\bigg(\ln \left(\frac{1}{a^2}\right) \Gamma \left(\frac{1}{2} \left(n+1\right),\frac{p}{a^2}\right)+G_{2,3}^{3,0}\left(\frac{p}{a^2}\left|
\begin{array}{c}
1,1 \\
0,0,\frac{n+1}{2} \\
\end{array}
\right.\right)\nonumber\\&+\Gamma \left(\frac{n+1}{2}\right) \left(\ln (p)-\psi ^{(0)}\left(\frac{n+1}{2}\right)\right)\bigg)
\label{eq:identity_3_asmyt_ber2}
\end{eqnarray}

\end{appendices}

\bibliographystyle{ieeetran}
\typeout{} 
\bibliography{tvt_bib}

\begin{thebibliography}{10}
\providecommand{\url}[1]{#1}
\csname url@samestyle\endcsname
\providecommand{\newblock}{\relax}
\providecommand{\bibinfo}[2]{#2}
\providecommand{\BIBentrySTDinterwordspacing}{\spaceskip=0pt\relax}
\providecommand{\BIBentryALTinterwordstretchfactor}{4}
\providecommand{\BIBentryALTinterwordspacing}{\spaceskip=\fontdimen2\font plus
\BIBentryALTinterwordstretchfactor\fontdimen3\font minus
  \fontdimen4\font\relax}
\providecommand{\BIBforeignlanguage}[2]{{%
\expandafter\ifx\csname l@#1\endcsname\relax
\typeout{** WARNING: IEEEtran.bst: No hyphenation pattern has been}%
\typeout{** loaded for the language `#1'. Using the pattern for}%
\typeout{** the default language instead.}%
\else
\language=\csname l@#1\endcsname
\fi
#2}}
\providecommand{\BIBdecl}{\relax}
\BIBdecl

\bibitem{Khalighi2014}
M.~A. {Khalighi} and M.~{Uysal}, ``Survey on free space optical communication:
  A communication theory perspective,'' \emph{IEEE Commun. Surveys Tuts.},
  vol.~16, no.~4, pp. 2231--2258, Fourthquarter 2014.

\bibitem{Bloom2003}
{S. Bloom \emph{et al.}}, ``Understanding the performance of free-space optics
  {[Invited]},'' \emph{J. Opt. Netw.}, vol.~2, no.~6, pp. 178--200, Jun 2003.

\bibitem{Kedar2004}
D.~{Kedar} and S.~{Arnon}, ``Urban optical wireless communication networks: the
  main challenges and possible solutions,'' \emph{IEEE Commun. Mag.}, vol.~42,
  no.~5, pp. S2--S7, May 2004.

\bibitem{Farid2007}
A.~A. {Farid} and S.~{Hranilovic}, ``Outage capacity optimization for
  free-space optical links with pointing errors,'' \emph{J. Lightw. Technol.},
  vol.~25, no.~7, pp. 1702--1710, July 2007.

\bibitem{Vavoulas2012}
{A. Vavoulas \emph{et al.}}, ``{Weather effects on {FSO} network
  connectivity},'' \emph{IEEE/OSA J. Opt. Commun. Netw.}, vol.~4, no.~10, pp.
  734--740, Oct 2012.

\bibitem{Kruse1962}
{P. W. Kruse \emph{et al.}}, \emph{Elements of Infrared Technology: Generation,
  Transmission and Detection}.\hskip 1em plus 0.5em minus 0.4em\relax New York:
  Wiley, 1962, vol.~1.

\bibitem{Kim2001}
{I. I. Kim \emph{et al.}}, ``Comparison of laser beam propagation at {785 nm}
  and {1550 nm} in fog and haze for optical wireless communications,''
  \emph{Proc. SPIE}, vol. 4214, pp. 26--37, Nov. 2001.

\bibitem{Bushuev2006}
D.~Bushuev and S.~Arnon, ``Analysis of the performance of a wireless optical
  multi-input to multi-output communication system,'' \emph{J. Opt. Soc. Am.
  A}, vol.~23, no.~7, pp. 1722--1730, Jul 2006.

\bibitem{Safari2008}
M.~{Safari} and M.~{Uysal}, ``Relay-assisted free-space optical
  communication,'' \emph{IEEE Trans. Wireless Commun.}, vol.~7, no.~12, pp.
  5441--5449, 2008.

\bibitem{Aghajanzadeh2011}
S.~M. {Aghajanzadeh} and M.~{Uysal}, ``Multi-hop coherent free-space optical
  communications over atmospheric turbulence channels,'' \emph{IEEE Trans.
  Commun.}, vol.~59, no.~6, pp. 1657--1663, 2011.

\bibitem{multi_hop_turb2015}
E.~{Zedini} and M.~{Alouini}, ``Multihop relaying over {IM/DD FSO} systems with
  pointing errors,'' \emph{J. Lightw. Technol.}, vol.~33, no.~23, pp.
  5007--5015, Dec 2015.

\bibitem{Chatzidiamantis2013}
{N. D. Chatzidiamantis \emph{et al.}}, ``Relay selection protocols for
  relay-assisted free-space optical systems,'' \emph{IEEE/OSA J. Opt. Commun.
  Netw.}, vol.~5, no.~1, pp. 92--103, 2013.

\bibitem{parallel_fso2015}
S.~{Molla Aghajanzadeh} and M.~{Uysal}, ``Performance analysis of parallel
  relaying in free-space optical systems,'' \emph{IEEE Trans. Commun.},
  vol.~63, no.~11, pp. 4314--4326, 2015.

\bibitem{multi_fso2013}
{M. A. Kashani \emph{et al.}}, ``Optimal relay placement and diversity analysis
  of relay-assisted free-space optical communication systems,'' \emph{IEEE/OSA
  J. Opt. Commun. Netw.}, vol.~5, no.~1, pp. 37--47, Jan 2013.

\bibitem{parallel_multi_fso2016}
C.~{Abou-Rjeily} and Z.~{Noun}, ``Impact of inter-relay co-operation on the
  performance of {FSO} systems with any number of relays,'' \emph{IEEE Trans.
  Wireless Commun.}, vol.~15, no.~6, pp. 3796--3809, June 2016.

\bibitem{Karimi2011}
M.~{Karimi} and M.~{Nasiri-Kenari}, ``Free space optical communications via
  optical amplify-and-forward relaying,'' \emph{J. Lightw. Technol.}, vol.~29,
  no.~2, pp. 242--248, 2011.

\bibitem{Kazemlou2011}
{Shabnam Kazemlou \emph{et al.}}, ``All-optical multihop free-space optical
  communication systems,'' \emph{J. Lightwave Technol.}, vol.~29, no.~18, pp.
  2663--2669, Sep 2011.

\bibitem{Bayaki2012}
E.~{Bayaki}, D.~S. {Michalopoulos}, and R.~{Schober}, ``{EDFA}-based
  all-optical relaying in free-space optical systems,'' \emph{IEEE Trans.
  Commun.}, vol.~60, no.~12, pp. 3797--3807, 2012.

\bibitem{Kashani2012}
{M. A. Kashani \emph{et al.}}, ``All-optical amplify-and-forward relaying
  system for atmospheric channels,'' \emph{IEEE Commun. Lett.}, vol.~16,
  no.~10, pp. 1684--1687, 2012.

\bibitem{Trinh2015}
{P. V. Trinh \emph{et al.}}, ``All-optical relaying {FSO} systems using {EDFA}
  combined with optical hard-limiter over atmospheric turbulence channels,''
  \emph{J. Lightw. Technol.}, vol.~33, no.~19, pp. 4132--4144, 2015.

\bibitem{Yang2014_relay}
{L. Yang \emph{et al.}}, ``Performance analysis of relay-assisted all-optical
  {FSO} networks over strong atmospheric turbulence channels with pointing
  errors,'' \emph{J. Lightw. Technol.}, vol.~32, no.~23, pp. 4613--4620, Dec
  2014.

\bibitem{dual_hop_turb2017}
{E. Zedini \emph{et al.}}, ``Dual hop {FSO} transmission systems over {Gamma
  Gamma} turbulence with pointing errors,'' \emph{IEEE Trans. Wireless
  Commun.}, vol.~16, no.~2, pp. 784--796, Feb 2017.

\bibitem{Dabiri2018}
M.~T. {Dabiri} and S.~M.~S. {Sadough}, ``Performance analysis of all-optical
  amplify and forward relaying over log-normal {FSO} channels,'' \emph{IEEE/OSA
  J. Opt. Commun. Netw.}, vol.~10, no.~2, pp. 79--89, 2018.

\bibitem{Huang2018}
{X. Huang \emph{et al.}}, ``Performance comparison of all-optical
  amplify-and-forward relaying {FSO} communication systems with {OOK} and
  {DPSK} modulations,'' \emph{IEEE Photon. J.}, vol.~10, no.~4, pp. 1--11,
  2018.

\bibitem{Khan2009}
{M. S. Khan \emph{et al.}}, ``Selecting a distribution function for optical
  attenuation in dense continental fog conditions,'' in \emph{2009 Int. Conf.
  on Emerg. Technol.}, 2009, pp. 142--147.

\bibitem{Esmail2016_Photonics}
{M. A. Esmail \emph{et al.}}, ``Outdoor {FSO} communications under fog:
  Attenuation modeling and performance evaluation,'' \emph{IEEE Photon. J.},
  vol.~8, no.~4, pp. 1--22, Aug 2016.

\bibitem{Esmail2017_Access}
------, ``On the performance of optical wireless links over random foggy
  channels,'' \emph{IEEE Access}, vol.~5, pp. 2894--2903, 2017.

\bibitem{Rahman2020}
{Z. Rahman \emph{et al.}}, ``Performance of opportunistic receiver beam
  selection in multiaperture {OWC} systems over foggy channels,'' \emph{IEEE
  Syst. J.}, vol.~14, no.~3, pp. 4036--4046, 2020.

\bibitem{Schimmel2018}
{G. Schimmel \emph{et al.}}, ``Free space laser telecommunication through
  fog,'' \emph{Optica}, vol.~5, no.~10, pp. 1338--1341, Oct 2018.

\bibitem{Esmail2017_Photonics}
{M. A. Esmail \emph{et al.}}, ``Outage probability analysis of {FSO} links over
  foggy channel,'' \emph{IEEE Photon. J.}, vol.~9, no.~2, pp. 1--12, 2017.

\bibitem{rahman2020cl}
{Z. Rahman \emph{et al.}}, ``Performance of opportunistic beam selection for
  {OWC} system under foggy channel with pointing error,'' \emph{IEEE Commun.
  Lett.}, vol.~24, no.~9, pp. 2029--2033, Sep. 2020.

\bibitem{Kedar2003}
D.~{Kedar} and S.~{Arnon}, ``Optical wireless communication through fog in the
  presence of pointing errors,'' \emph{Appl. Opt.}, vol.~42, no.~24, pp.
  4946--4954, Aug. 2003.

\bibitem{Naboulsi2004}
{M. Al Naboulsi}, ``Fog attenuation prediction for optical and infrared
  waves,'' \emph{Opt. Eng.}, vol.~43, no.~2, pp. 319--329, 2004.

\bibitem{Awan2008}
{M. Saleem Awan \emph{et al.}}, ``Attenuation analysis for optical wireless
  link measurements under moderate continental fog conditions at {Milan and
  Graz},'' in \emph{IEEE 68th Veh. Technol. Conf. (VTC) 2008}, 2008, pp. 1--5.

\bibitem{Esmail2016_ICC}
{M. A. Esmail \emph{et al.}}, ``Analysis of fog effects on terrestrial free
  space optical communication links,'' in \emph{2016 IEEE Int. Conf. on Commun.
  Wkshp. (ICC)}, May 2016, pp. 151--156.

\bibitem{Berenguer2018}
{P. Wilke Berenguer \emph{et al.}}, ``Optical wireless {MIMO} experiments in an
  industrial environment,'' \emph{IEEE J. Sel. Areas Commun.}, vol.~36, no.~1,
  pp. 185--193, Jan 2018.

\bibitem{Zhu2002}
X.~{Zhu} and J.~M. {Kahn}, ``Free-space optical communication through
  atmospheric turbulence channels,'' \emph{IEEE Trans. Commun.}, vol.~50,
  no.~8, pp. 1293--1300, 2002.

\bibitem{Barrios2012}
R.~Barrios and F.~Dios, ``Exponentiated {Weibull} distribution family under
  aperture averaging for {Gaussian} beam waves,'' \emph{Optics Express},
  vol.~20, no.~12, pp. 13\,055--13\,064, Jun 2012.

\bibitem{Andrews20015}
L.~C. Andrews and R.~L. Phillips, \emph{Laser Beam Propagation Through Random
  Media, vol. 1. Bellingham}.\hskip 1em plus 0.5em minus 0.4em\relax SPIE,
  2005, vol.~1.

\bibitem{malaga2011}
A.~Jurado-Navas \emph{et~al.}, ``A unifying statistical model for atmospheric
  optical scintillation,'' \emph{Numerical Simulations of Physical and
  Engineering Processes}, Sep 2011.

\bibitem{Peppas2020}
{K. P. {Peppas} \emph{et al.}}, ``The fischer–snedecor $\mathcal
  {F}$-distribution model for turbulence-induced fading in free-space optical
  systems,'' \emph{J. Lightw. Technol.}, vol.~38, no.~6, pp. 1286--1295, 2020.

\bibitem{Kashani2015}
M.~A. Kashani, M.~Uysal, and M.~Kavehrad, ``A novel statistical model for
  turbulence-induced fading in free-space optical systems,'' in \emph{2013 15th
  International Conference on Transparent Optical Networks (ICTON)}, 2013, pp.
  1--5.

\bibitem{AlQuwaiee2015}
H.~AlQuwaiee \emph{et~al.}, ``On the performance of free-space optical
  communication systems over double {Generalized Gamma} channel,'' \emph{IEEE
  Journal on Selected Areas in Communications}, vol.~33, no.~9, pp. 1829--1840,
  2015.

\bibitem{Ashrafzadeh2020}
B.~Ashrafzadeh, A.~Zaimbashi, E.~Soleimani-Nasab, and M.~Uysal, ``Unified
  performance analysis of multi-hop {FSO} systems over double generalized gamma
  turbulence channels with pointing errors,'' \emph{IEEE Transactions on
  Wireless Communications}, vol.~19, no.~11, pp. 7732--7746, 2020.

\bibitem{Nosratinia2004}
{A. Nosratinia \emph{et al.}}, ``Cooperative communication in wireless
  networks,'' \emph{IEEE Commun. Mag.}, vol.~42, no.~10, pp. 74--80, 2004.

\bibitem{Li2012}
{Q. Li \emph{et al.}}, ``Cooperative communications for wireless networks:
  techniques and applications in {LTE}-advanced systems,'' \emph{IEEE Wireless
  Commun.}, vol.~19, no.~2, pp. 22--29, 2012.

\bibitem{Lee2011}
{E. Lee \emph{et al.}}, ``Performance analysis of the asymmetric dual-hop relay
  transmission with mixed {RF/FSO} links,'' \emph{IEEE Photon. Technol. Lett.},
  vol.~23, no.~21, pp. 1642--1644, 2011.

\bibitem{Ansari2013}
{I. S. Ansari \emph{et al.}}, ``Impact of pointing errors on the performance of
  mixed {RF/FSO} dual-hop transmission systems,'' \emph{IEEE Wireless Commun.
  Lett.}, vol.~2, no.~3, pp. 351--354, 2013.

\bibitem{Samimi2013}
H.~{Samimi} and M.~{Uysal}, ``End-to-end performance of mixed {RF/FSO}
  transmission systems,'' \emph{IEEE/OSA J. Opt. Commun. Netw.}, vol.~5,
  no.~11, pp. 1139--1144, 2013.

\bibitem{assym_rf_fso2015}
S.~{Anees} and M.~R. {Bhatnagar}, ``Performance of an {amplify-and-forward}
  dual hop asymmetric {RF-FSO} communication system,'' \emph{IEEE/OSA J. Opt.
  Commun. Netw.}, vol.~7, no.~2, pp. 124--135, February 2015.

\bibitem{series_hybrid_m_channel2015}
{L. Kong \emph{et al.}}, ``Performance of a free-space-optical relay-assisted
  hybrid {RF/FSO} system in generalized {$M$}-distributed channels,''
  \emph{IEEE Photon. J.}, vol.~7, no.~5, pp. 1--19, Oct 2015.

\bibitem{dual_hop_rf_fso_turb2016}
{E. Zedini \emph{et al.}}, ``On the performance analysis of dual-hop mixed
  {FSO/RF} systems,'' \emph{IEEE Trans. Wireless Commun.}, vol.~15, no.~5, pp.
  3679--3689, May 2016.

\bibitem{Bag2018}
{B. Bag \emph{et al.}}, ``Performance analysis of hybrid {FSO} systems using
  {FSO/RF-FSO} link adaptation,'' \emph{IEEE Photon. J.}, vol.~10, no.~3, pp.
  1--17, 2018.

\bibitem{Zhang2020}
{Y. Zhang \emph{et al.}}, ``On the performance of dual-hop systems over mixed
  {FSO/mmWave} fading channels,'' \emph{IEEE Open J. Commun. Soc.}, vol.~1, pp.
  477--489, 2020.

\bibitem{Papoulis2001}
A.~Papoulis and U.~Pillai, \emph{\BIBforeignlanguage{English (US)}{Probability,
  random variables and stochastic processes}}, 4th~ed.\hskip 1em plus 0.5em
  minus 0.4em\relax McGraw-Hill, Nov. 2001.

\bibitem{Kilbas}
A.~{Kilbas}, \emph{Analytical methods and special functions, {H-Transforms}
  Theory and Applications}.\hskip 1em plus 0.5em minus 0.4em\relax New York,
  NY, USA: Taylor and Francis, 2004.

\bibitem{Zwillinger2014}
{ D. {Zwillinger}}, \emph{Table of integrals, series, and products}.\hskip 1em
  plus 0.5em minus 0.4em\relax Elsevier, 2014.

\bibitem{Wolfram}
\BIBentryALTinterwordspacing
{Wolfram. (2001) The Wolfram functions site. Internet.} [Online]. Available:
  \url{http://functions.wolfram.com}
\BIBentrySTDinterwordspacing

\bibitem{Tsiftsis2006}
T.~A. Tsiftsis, H.~G. Sandalidis, G.~K. Karagiannidis, and N.~C. Sagias,
  ``Multihop free-space optical communications over strong turbulence
  channels,'' in \emph{2006 IEEE International Conference on Communications},
  vol.~6, 2006, pp. 2755--2759.

\end{thebibliography}

\begin{IEEEbiography}[{\includegraphics[width=1.1in,height=1.35in,clip,keepaspectratio]{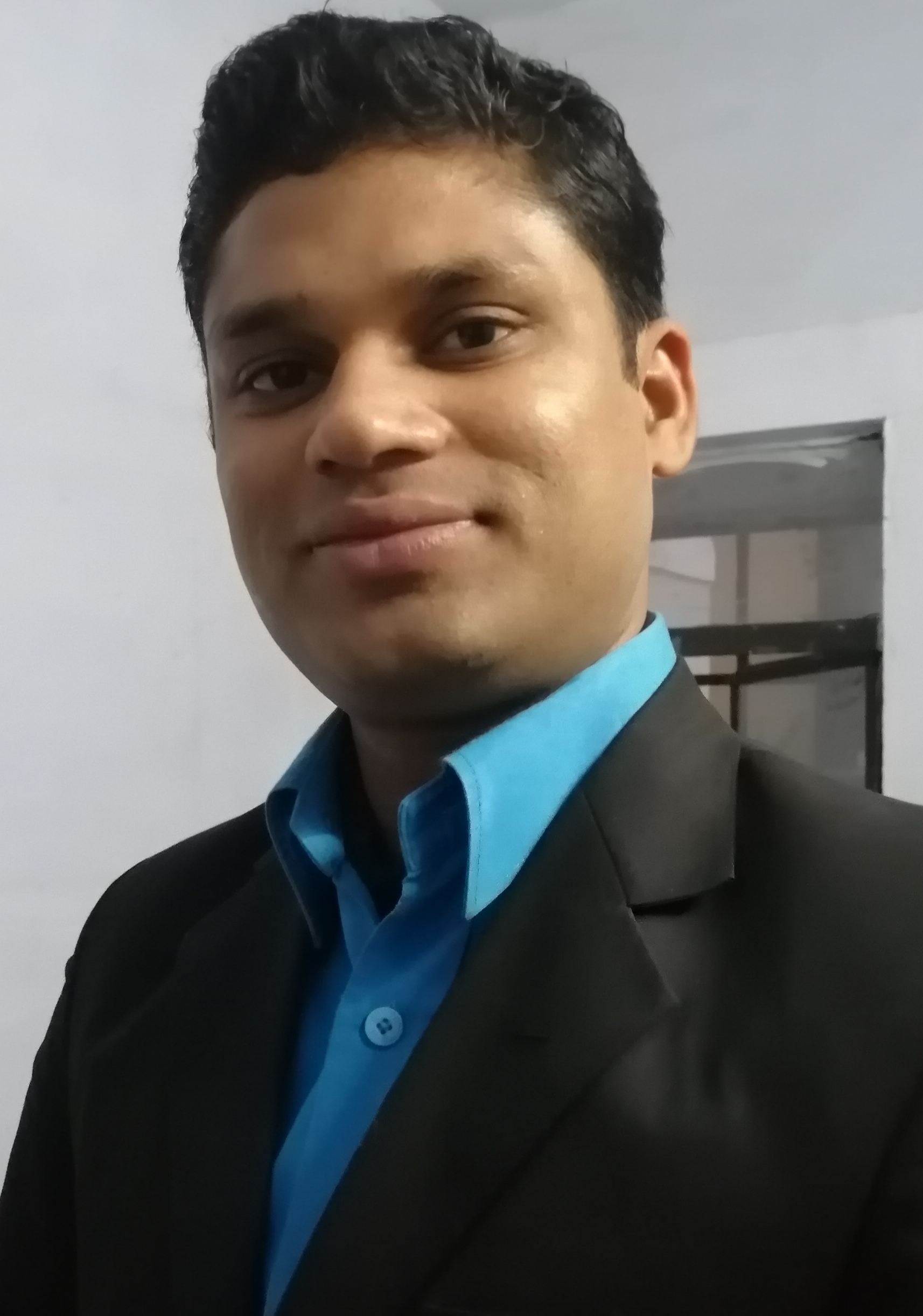}}]%
	{Ziyaur Rahman} (Student Member, IEEE)  received the B.Tech. and the M.Tech. degrees in electronics and communication engineering from Punjab Technical University, Kapurthala, India, in 2015 and 2017, respectively. He is currently pursuing the Ph.D. degree in the area of wireless communications with the Department of Electrical and Electronics Engineering, Birla Institute of Technology and Science at Pilani, Pilani, India.
	His  current research interests include optical wireless communications for terrestrial applications and machine learning for communication systems.
		
\end{IEEEbiography}
\begin{IEEEbiography}[{\includegraphics[width=1.1in,height=1.35in,clip,keepaspectratio]{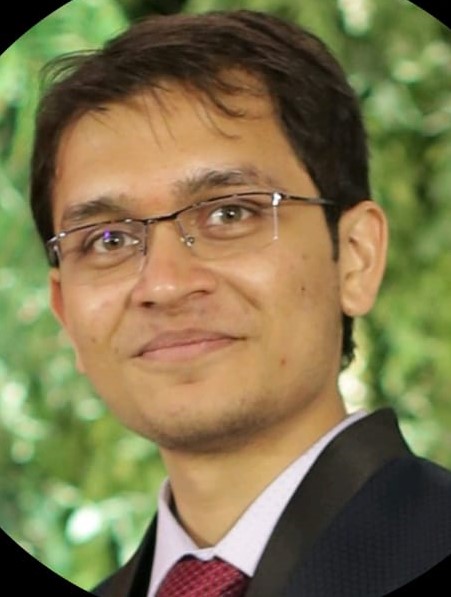}}]%
	{Tejas Nimish Shah}  received the B.Tech. degree in Electronics and Electrical Engineering from Birla Institute of Technology and Science at Pilani, Pilani, India, in 2021. Currently, he is working as a Risk Management Analyst in Nomura Capital (India) Private Limited, Mumbai, India. He has completed research projects on  optical wireless communications, adaptive feed control, and automated tuning of PID controllers. His current research interests include algorithmic finance and investment,  IoT, and wireless communications.		
\end{IEEEbiography}

\begin{IEEEbiography}[{\includegraphics[width=1.1in,height=1.35in,clip,keepaspectratio]{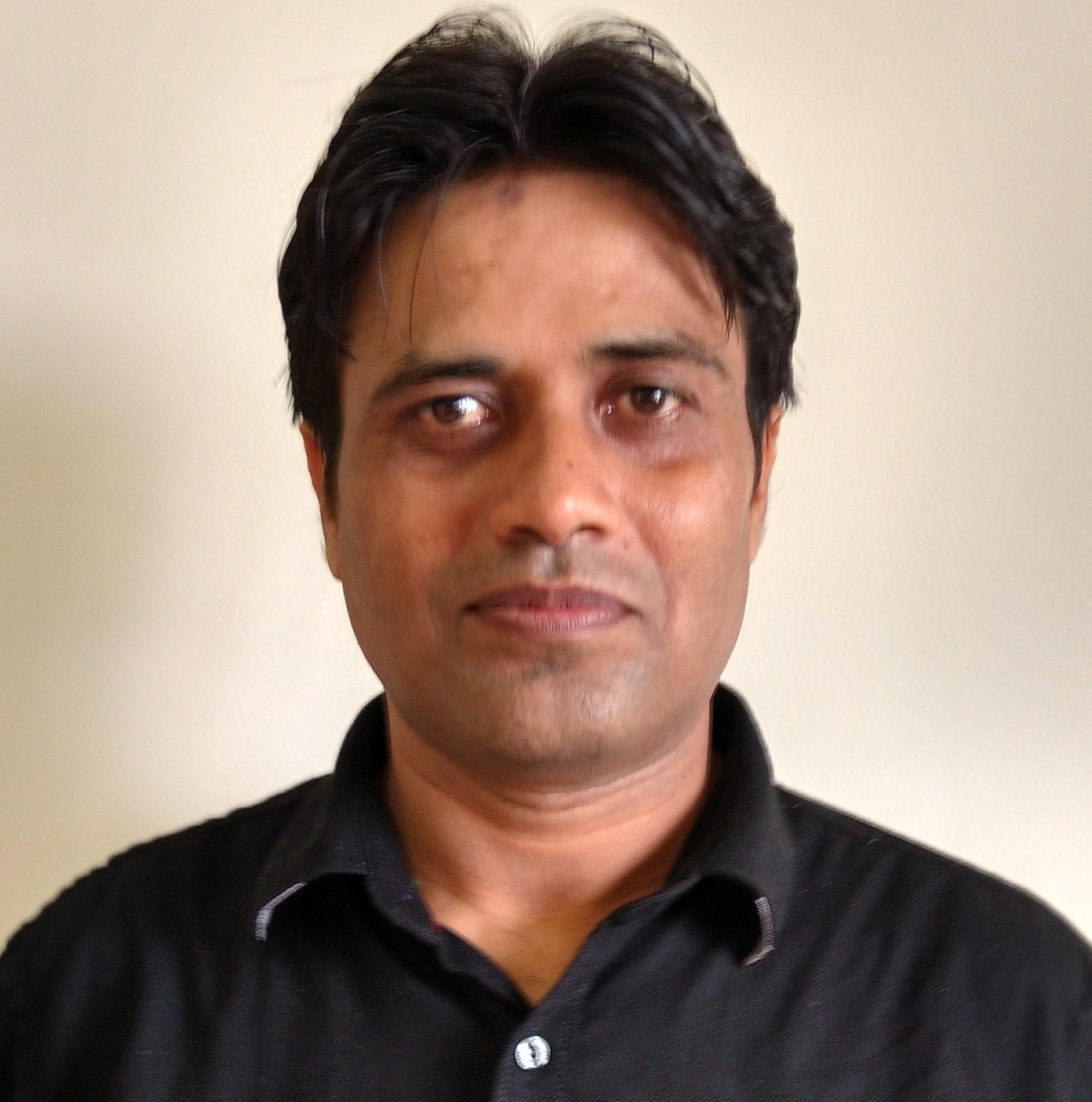}}]%
	{S.~M.~Zafaruddin } (Senior Member, IEEE) received the Ph.D.  degree in electrical engineering from IIT Delhi,  New Delhi, India, in 2013. From 2012 to 2015, he was with Ikanos Communications (now Qualcomm), Bengaluru, India, working directly with  the CTO Office, Red Bank, NJ, USA, where he  was involved in research and development for  xDSL systems. From 2015 to 2018, he was a Post-Doctoral Researcher with the Faculty of Engineering, Bar-Ilan University, Ramat Gan, Israel, where he was involved in signal processing for  wireline and wireless communications. He is currently an Assistant Professor with  the Department of Electrical and Electronics Engineering, Birla Institute of Technology and Science at Pilani, Pilani, India. His current research interests  include   signal processing and machine learning for wireless and wireline  communications, THz wireless technology, optical wireless communications, reconfigurable intelligent surface, distributed signal processing,  and resource allocation algorithms. He received the Planning and Budgeting Commission Fellowship for  Outstanding Post-Doctoral Researchers from China and India by the Council  for Higher Education, Israel (2016–2018). He is also an Associate Editor of the IEEE ACCESS.
\end{IEEEbiography}

\begin{IEEEbiography}[{\includegraphics[width=1.1in,height=1.35in,clip,keepaspectratio]{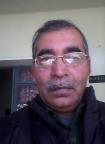}}]%
	{Vinod Kumar Chaubey } (Senior Member, IEEE) received the master's degree with specialization in electronics and radio physics and the Ph.D. degree in the field of fiber optics communication from Banaras Hindu University (BHU), Varanasi, India, in 1985 and 1992, respectively.
	He was with the Department of Applied Physics, Institute of Technology, BHU, during 1993 and 1994 as a UGC Postdoctoral Fellow. He joined the Birla Institute of Technology \& Science at Pilani, Pilani, India, in 1994 as a Lecturer and is currently working as a Professor (since 2010) with the Department of Electrical \& Electronic Engineering. His research interests include wireless and optical communication network design, electronic circuits design, and modeling and simulation.
	Dr. Chaubey is an IETE Fellow.
\end{IEEEbiography}

\end{document}